\documentclass[usenatbib]{mnras}
\usepackage{newtxtext,newtxmath}
\usepackage[T1]{fontenc}
\DeclareRobustCommand{\VAN}[3]{#2}
\let\VANthebibliography\thebibliography
\def\thebibliography{\DeclareRobustCommand{\VAN}[3]{##3}\VANthebibliography}

\usepackage{graphicx}	
\usepackage{amsmath}	
\newcommand{\Msun}{M$_\odot$} 

\title[Galaxy quenching by SMBHs in FIRE simulations]{Exploring supermassive black hole physics and galaxy quenching across halo mass in FIRE cosmological zoom simulations}

\author[S. Wellons et al.]{Sarah Wellons$^{1}$\thanks{E-mail: sarah.wellons@northwestern.edu}, Claude-Andr\'e Faucher-Gigu\`ere$^{1}$, Philip F. Hopkins$^{2}$, Eliot Quataert$^{3}$, \newauthor
Daniel Angl\'es-Alc\'azar$^{4,5}$, Robert Feldmann$^{6}$, Christopher C. Hayward$^{5}$, Du\v{s}an Kere\v{s}$^{7}$, \newauthor
Kung-Yi Su$^{5,8}$, and Andrew Wetzel$^{9}$
\\
$^{1}$CIERA and Department of Physics and Astronomy, Northwestern University, 1800 Sherman Ave, Evanston, IL 60201, USA \\
$^{2}$TAPIR, MC 350-17, California Institute of Technology, Pasadena, CA 91125, USA \\
$^{3}$Department of Astrophysical Sciences, Princeton University, Princeton, NJ 08544, USA \\
$^{4}$Department of Physics, University of Connecticut, 196 Auditorium Road, U-3046, Storrs, CT 06269-3046, USA \\
$^{5}$Center for Computational Astrophysics, Flatiron Institute, 162 Fifth Avenue, New York, NY 10010, USA \\
$^{6}$Institute for Computational Science, University of Zurich, Zurich CH-8057, Switzerland \\
$^{7}$Department of Physics, Center for Astrophysics and Space Sciences, University of California, San Diego, La Jolla, CA, USA \\
$^{8}$Department of Astronomy, Columbia University, 550 West 120th Street, New York, NY 10027, USA \\
$^{9}$Department of Physics and Astronomy, University of California, Davis, CA 95616, USA
}

\date{Accepted XXX. Received YYY; in original form ZZZ}
\pubyear{2020}

\begin{document}
\label{firstpage}
\pagerange{\pageref{firstpage}--\pageref{lastpage}}
\maketitle

\begin{abstract}
Feedback from accreting supermassive black holes (SMBHs) is thought to be a primary driver of quenching in massive galaxies, but the best way to implement SMBH physics into galaxy formation simulations remains ambiguous.  As part of the Feedback in Realistic Environments (FIRE) project, we explore the effects of different modeling choices for SMBH accretion and feedback in a suite of $\sim500$ cosmological zoom-in simulations across a wide range of halo mass ($10^{10}-10^{13}$ \Msun). Within the suite, we vary the numerical schemes for BH accretion and feedback, the accretion efficiency, and the strength of mechanical, radiative, and cosmic ray feedback independently. We then compare the outcomes to observed galaxy scaling relations. We find several models that satisfy the observational constraints, and for which the energetics in different feedback channels are physically plausible. Interestingly, cosmic rays accelerated by SMBHs play an important role in many successful models. However, it is non-trivial to reproduce scaling relations across halo mass, and many model variations produce qualitatively incorrect results regardless of parameter choices. The growth of stellar and BH mass are closely related: for example, over-massive BHs tend to over-quench galaxies. BH mass is most strongly affected by the choice of accretion efficiency in high-mass halos, but by feedback efficiency in low-mass halos.  The amount of star formation suppression by SMBH feedback in low-mass halos is determined primarily by the time-integrated feedback energy.  For massive galaxies, the ``responsiveness" of a model (i.e. how quickly and powerfully the BH responds to gas available for accretion) is an additional important factor for quenching.
\end{abstract}

\begin{keywords}
galaxies: active -- galaxies: evolution -- galaxies: formation -- galaxies: quasars: general -- galaxies: quasars: supermassive black holes -- galaxies: star formation
\end{keywords}



\section{Introduction}

The question of how galaxies ``quench" (i.e., stop actively forming new stars) is one of the fundamental outstanding problems in galaxy formation.  The observational evidence is clear that above a stellar mass of approximately $10^{10.5}$ \Msun, galaxies begin to cease forming stars and become ``red and dead" even in the absence of the environmental effects (e.g., ram pressure stripping) that are known to induce quenching in lower-mass systems, and that this population of massive red galaxies builds up continuously over cosmic time \citep{Bell2003, Kauffmann2003, Baldry2004, Blanton2005, Faber2007, Peng2010a, Pozzetti2010, Muzzin2013, Tinker2013, Tomczak2014}.  Some physical process is acting in these massive systems to prevent gas from cooling, condensing, and forming stars.  

Many potential quenching mechanisms have been proposed and discussed in the literature, such as the shock-heating of gas as it falls into massive halos \citep{Dekel2006a, Keres2009}, magnetic fields and/or thermal conduction in the circumgalactic medium \citep{Beck2012, Wagh2014, Dolag2004, Voit2011, Parrish2012}, and the alteration of the gravitational potential by changes in galaxy morphology \citep{Martig2009, Dekel2009a}.  

Another promising possibility (which is the main focus of our study) is that the highly energetic winds, jets, and radiation driven by accretion onto supermassive black holes (SMBHs) impact the gas supply into their host galaxies and contribute to the observed quenching in massive systems.  In general, previous simulation studies have found that massive galaxies fail to quench to the degree seen in observations when SMBH (or ``active galactic nucleus," AGN) accretion and feedback is not included, even when some or all of the other aforementioned mechanisms are present \citep{Sijacki2007, Booth2009, Choi2015, Eisenreich2017, Feldmann2017, Su2019}. As a result, most modern simulations of galaxy formation which seek to reproduce the known observational statistics incorporate some model for accretion onto and feedback from SMBHs to induce a population of red galaxies at the massive end \citep{Somerville2015}.  

Although great progress has been made in the direct simulation of BH accretion disks, winds, and jets (see e.g. the review by \citet{Davis2021}), this physics occurs on scales well below the resolution currently achievable in cosmological simulations of galaxy formation.  Thus, any representation of SMBH physics in these simulations must rely on "sub-grid" modeling of the inflow to the black hole accretion disk system and the emergent outflows at distances of several pc (or greater, depending on resolution).   

Because the fundamental physics is unresolved, the details of how black holes are modeled vary considerably between galaxy formation simulations.  In the original Illustris simulation, a large-volume cosmological simulation with a box size of (100 Mpc)$^3$, black hole accretion was modeled using an Eddington-limited \citet{Bondi1944} prescription, and feedback was injected thermally (at high accretion rates or ``quasar mode") or in radio bubbles \citep[at low accretion rates or ``radio mode";][]{Sijacki2007, Sijacki2015, Vogelsberger2013}.  In its successor IllustrisTNG, the low-accretion-rate mode was modified to a kinetic, stochastically-generated wind \citep{Weinberger2017, Pillepich2018}.  The EAGLE project uses a single-mode AGN feedback model with stochastic, thermal injections proportional to the BH accretion rate, similarly given by an Eddington-limited Bondi model with an additional consideration of the angular momentum of the gas \citep{Rosas-Guevara2015,Schaye2015,Crain2015}.  In the SIMBA simulations, accretion is modeled in two modes, with the accretion of cold gas driven by gravitational torques and the accretion of hot gas described by a Bondi model, while feedback is modeled kinetically with wind speeds dependent on accretion rate \citep{Angles-Alcazar2017b,Dave2019}.  Each of these representations is physically sensible, but also physically distinct - these varying implementations leave varying imprints on the resulting galaxies, and it can be unclear what constitutes a meaningful prediction and what is simply a property of that particular model.

Some observational constraints do exist to help narrow down which galaxy-scale implementations of black hole physics are the most physically plausible \citep[e.g.,][]{moe09:bal.energetics,fg12:felobals,stern16:agn.mechanisms,richings21:emission.tracers}. 
AGN can be highly luminous in X-rays, optical, or infrared emission, and their intense radiation pressure has been shown to drive strong winds in multiple phases of gas which inject a significant amount of energy and momentum back into the galaxy \citep[e.g.,][]{fg12:energy.conserving,zubovas12:clearing, costa14:energy.momentum,richings18:sims,richings18:analytic}. 
Radio jets emanating from SMBHs have been observed to carve out huge cavities and drive shock fronts in the circumgalactic medium of galaxy clusters \citep{McNamara2012}.  In recent years, AGN have also been discovered to be acting in lower-mass ($M_* < 10^{10}$ \Msun) systems \citep{Penny2018, Bradford2018, Dickey2019}. 
For more on observational constraints on AGN feedback, see the reviews by \cite{fabian12:review}, \citet{Harrison2017} and \citet{Morganti2017}. 

Despite the clear observational evidence for the action of AGN on galaxies through radiation, winds, and jets, in detail there remains a vast parameter space of physical plausibility, making it difficult to define a black hole physics model {\it a priori} for use in simulations.  The goal of this study is to explore this parameter space and determine the common attributes of models which lead to quenching behavior in line with known observational statistics across a broad range of halo mass.

In contrast to the simulation works described above, we employ a zoom-in technique to simulate individual halos at higher resolution than is available in large-volume simulations, while preserving the cosmological context.  As part of the Feedback In Realistic Environments (FIRE)\footnote{See the FIRE project web site: http://fire.northwestern.edu.} project \citep{Hopkins2014}, we use the FIRE-2 galaxy formation model \citep{Hopkins2018} to model the physics of stars and the interstellar medium (ISM).  Because the relatively high resolution allows us to capture the multiphase nature of the ISM, we are able to explicitly describe physics which is not included in other simulations (such as the interaction between AGN feedback and a realistic multiphase ISM).  Other studies using the FIRE model for ISM and stellar physics have explored quenching behavior in idealized simulations of disk galaxies \citep{Torrey2020} and very massive ($>10^{14}$ \Msun) halos \citep{Su2020, Su2021}, with black holes accreting at a fixed rate, or have investigated the mechanisms driving black hole growth in the absence of AGN feedback \citep{Angles-Alcazar2017, Angles-Alcazar2021}.
In this study, we examine a suite of zoom-in simulations of systems with halo masses from $10^{11}-10^{13}$ \Msun, run down to $z=0$ (for halos $<10^{12}$ \Msun) or $z=1$ (for halos $>10^{12}$ \Msun), with ``live" BH accretion that is actively influenced by stellar and BH feedback.

We start from a position of trying to represent the physics as faithfully as possible by implementing all known feedback channels through which SMBHs interact with the surrounding gas, including mechanical feedback, radiation pressure, and cosmic rays.  We represent uncertainty in the unresolved physics at the BH scale with several parameters which we vary within the range permitted by observations.  Rather than trying to identify a single ``best" model for black hole physics, we explore the parameter space to determine the overall properties of models that lead to quenching in high-mass systems, and those that avoid quenching in low-mass systems, to gain insight about the underlying physics.

In Section \ref{sec:methods}, we describe the suite of zoom-in simulations run and analyzed for this study, the various models for SMBH accretion and feedback employed, and the parameter space explored.  In Section \ref{sec:results}, we present the outcomes of the simulations in terms of the regulation of BH and stellar mass growth and quenching, and show several illustrative examples of models which succeed or fail according to known observational scaling relations.  In Section~\ref{sec:bhreg}, we demonstrate the relationship between BH growth and stellar mass growth and show how the choices of feedback and accretion efficiencies affect BH mass.  In Section~\ref{sec:sfreg}, we discuss how the amount of feedback energy and the responsiveness of the SMBH model can affect the regulation of star formation in galaxies.  In Section~\ref{sec:goodmodels} we show examples of models which perform well across halo mass and discuss their properties and behaviors.  We summarize our findings and conclude in Section \ref{sec:conclusions}.

\section{Simulations and Methods}
\label{sec:methods}

\subsection{Simulations}
\label{ssec:sims} 

All simulations analyzed in this work are cosmological zoom-in simulations which are part of the Feedback In Realistic Environments (FIRE) project \citep{Hopkins2014}, evolved using the GIZMO\footnote{http://www.tapir.caltech.edu/$\sim$phopkins/Site/GIZMO.html} meshless finite mass (MFM) hydrodynamics solver \citep{Hopkins2015} and the FIRE-2 galaxy formation model.  We include all the ``standard" FIRE-2 physics (i.e., gas cooling down to 10 K, star formation in dense, self-gravitating gas, and stellar feedback driven by radiation pressure, photo-ionization, photo-electric heating, stellar winds, and supernovae) described in detail by \citet{Hopkins2018}. In addition, we include updated stellar feedback from cosmic rays (CRs) as described by \citet{Chan2019}, where $\sim 10\%$ of the initial ejecta energy from SNe and fast stellar winds is injected as CRs, for which we explicitly evolve the two-moment (flux+energy) equations including streaming at the Alfven speed plus a constant cosmic ray scattering rate equivalent to a constant parallel (anisotropic) diffusion coefficient $\kappa_{\|} \approx 3\times10^{29}\,{\rm cm^{2}\,s^{-1}}$ (calibrated to reproduce detailed Solar-system observations of CRs as well as $\gamma$-ray observations of nearby galaxies; see \citealt{Chan2019,hopkins:cr.transport.constraints.from.galaxies,hopkins:cr.multibin.mw.comparison}).

The simulation suite includes 21 different halos with masses ranging from $10^{10} - 10^{13}$ \Msun, with mass resolution scaling with halo mass from 250 \Msun~to $2.6 \times 10^{5}$ \Msun~as shown in Table~\ref{tab:sims}.  The most massive halos (whose labels begin with `m13') were run only to redshift $z=1$ due to their computational expense (as the $z=0$ zoom-in region would be much larger); all other simulations were continued to $z=0$.   Each halo was run multiple times with variations of the SMBH accretion and feedback models described in the following subsections.  The greatest number of variations were run on one halo per decade of mass: m11i ($M_{\rm h} = 6.7 \times 10^{10}$ \Msun~at $z=0$), m12i ($M_{\rm h} = 10^{12}$ \Msun~at $z=0$), and m13A1 ($M_{\rm h} = 4.3 \times 10^{12}$ \Msun~at $z=1$).  The following subsections describe the elements of SMBH physics modeled in this work.  (For additional implementation details, see the FIRE-3 methods paper, \citet{hopkins:fire3.methods}.)

\subsection{Black hole seeding and dynamics}
\label{ssec:bhinit}

Black hole seed particles of 100 \Msun~are generated probabilistically from gas cells which meet the criteria for star formation and are not near another BH particle. The cell is assigned a probability of forming a BH seed based on its surface density and metallicity, with strong weighting towards high density and low metallicity. The probability function falls off sharply at surface densities below 5000 \Msun/pc$^2$ and metallicities above 0.001 $Z_\odot$.  Once formed, BH particles move according to the normal laws of gravity, plus an additional artificial acceleration term towards the most-bound collisionless particle in its interaction kernel each timestep. This term has the effect of moving the BH particle smoothly towards the galaxy center. BH particles whose interaction kernels overlap and are mutually bound are merged instantaneously. We note that the question of how SMBHs actually move and merge in realistic galactic potentials is an interesting one (see e.g. \citet{Ma2021} and the Romulus simulations by \citet{Tremmel2017, Tremmel2019}), but we defer its study to future work as we focus here on accretion and feedback from black holes assumed to be located in galaxy centers.

In a preliminary set of simulations, we have systematically varied the seeding criteria, seed masses, ``drag'' term, and merger criteria above. For the range of models studied here, we find that all our results are insensitive to these choices so long as the seeding algorithm and BH dynamics allow for each galaxy to have at least $\sim1$ seed which can remain near the galaxy center (although we stress this is not trivial to achieve, in practice).

\begin{figure*}
  \centering
  \includegraphics[width=2\columnwidth]{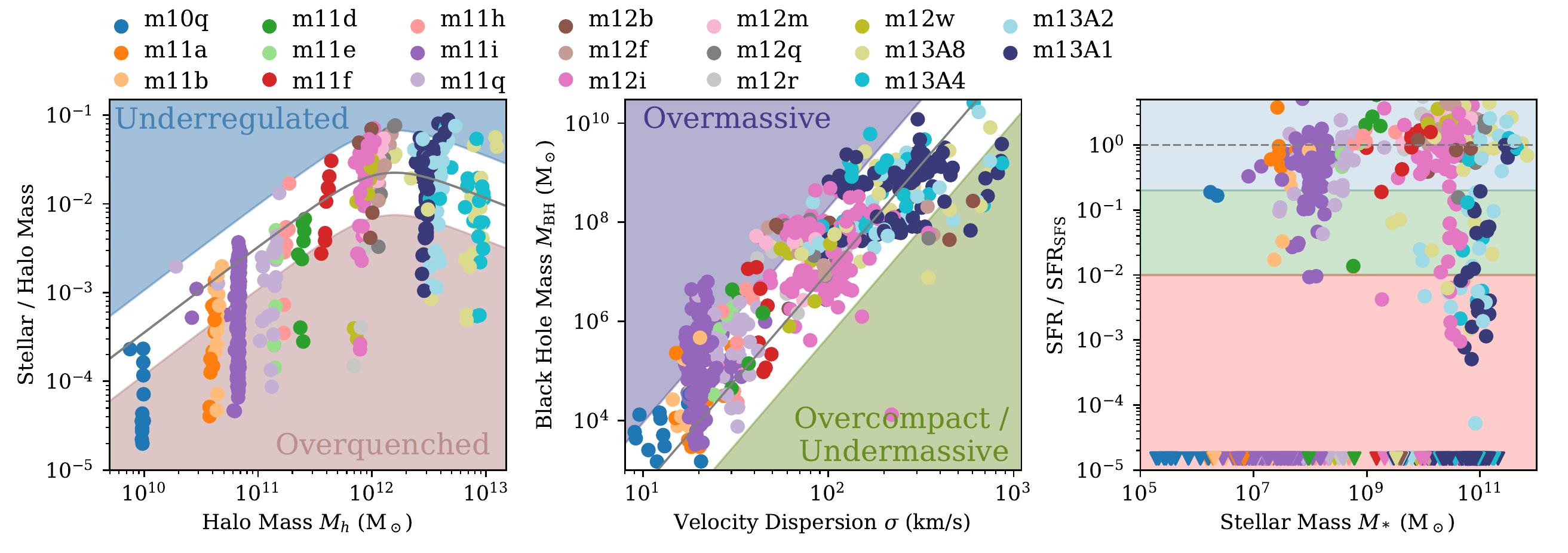}
  \caption{All simulations in our AGN suite at their final snapshot ($z=1$ for m13X simulations, $z=0$ for all others).  {\it Left:} Stellar mass - halo mass relation, indicating the level of star formation regulation in the galaxy.  The $z=0$ curve from \citet{Behroozi2019} is shown in grey for comparison.  Many implementations of AGN physics over-quench galaxies across halo mass, some fail to quench massive galaxies, and some lead to an appropriate level of star formation regulation. {\it Center:} The $M_{\rm BH}$-$\sigma$ relation, indicating the amount of black hole growth relative to the galaxy.  The observed relation from \citet{Greene2020} is shown for comparison.  {\it Right:}  Star formation rate in the galaxy, relative to the median value of the star-forming sequence for that stellar mass and redshift from \citet{Whitaker2014} and \citet{Salim2007}, modified as suggested by \citet{Leja2019}.  Triangles at the bottom of the figure indicate star formation rates below the range plotted.}
  \label{fig:allsims}
\end{figure*}

\begin{table*}
\begin{center}
\begin{tabular}{|c|cccc|cccc|} 
 \hline
 Halo Name & Resolution (\Msun) & $z_{\rm final}$ & $M_{\rm halo}$ at $z_{\rm final}$ & N$_{\rm runs}$ & $\eta_{\rm CR}$ & $v_{\rm wind}$ (km/s) & $\eta_{\rm RP}$ & $\bar{\eta}_{\rm acc}$ \\ 
 \hline\hline
 m13A8 & $2.6 \times 10^5$ & 1 & $1.2 \times 10^{13}$ & 25 & $0.0001 - 0.01$ & $1000 - 30000$ & $0.1 - 1$ & $0.007 - 9$ \\ 
 m13A2 & $2.6 \times 10^5$ & 1 & $8.1 \times 10^{12}$ & 25 & $0.0001 - 0.01$ & $1000 - 30000$ & $0.1 - 1$ & $0.007 - 9$ \\ 
 m13A4 & $2.6 \times 10^5$ & 1 & $4.7 \times 10^{12}$ & 27 & $0 - 0.01$ & $0 - 42500$ & $0 - 100$ & $0.007 - 9$ \\ 
 \hline
 \textbf{m13A1} & $2.6 \times 10^5$ & 1 & $4.3 \times 10^{12}$ & 109 & $0 - 0.1$ & $0 - 42500$ & $0 - 100$ & $0.003 - 12$ \\ 
 \hline
 m12b & $5.7 \times 10^4$ & 0 & $1.1 \times 10^{12}$ & 5 & $0.001 - 0.01$ & $3000 - 10000$ & $0.1 - 10$ & $0.012 - 1.8$ \\ 
 m12f & $5.7 \times 10^4$ & 0 & $1.4 \times 10^{12}$ & 7 & $0.001 - 0.01$ & $3000 - 10000$ & $0.1 - 10$ & $0.007 - 1.8$ \\ 
 \hline
 \textbf{m12i} & $5.7 \times 10^4$ & 0 & $1 \times 10^{12}$ & 81 & $0 - 0.1$ & $0 - 42500$ & $0 - 100$ & $0.002 - 12$ \\ 
 \hline
 m12m & $5.7 \times 10^4$ & 0 & $1.3 \times 10^{12}$ & 5 & $0.001 - 0.01$ & $3000 - 10000$ & $0.1 - 10$ & $0.012 - 1.8$ \\ 
 m12q & $5.7 \times 10^4$ & 0 & $1.6 \times 10^{12}$ & 7 & $0.001 - 0.01$ & $3000 - 10000$ & $0.1 - 10$ & $0.007 - 1.8$ \\ 
 m12r & $5.7 \times 10^4$ & 0 & $9.2 \times 10^{11}$ & 5 & $0.001 - 0.01$ & $3000 - 10000$ & $0.1 - 10$ & $0.012 - 1.8$ \\ 
 m12w & $5.7 \times 10^4$ & 0 & $9.5 \times 10^{12}$ & 5 & $0.001 - 0.01$ & $3000 - 10000$ & $0.1 - 10$ & $0.012 - 1.8$ \\ 
 m11a & 2100 & 0 & $4.2 \times 10^{10}$ & 9 & $0.001 - 0.01$ & $3000 - 10000$ & $0.1 - 1$ & $0.007 - 1.8$ \\ 
 m11b & 2100 & 0 & $4.8 \times 10^{10}$ & 9 & $0.001 - 0.01$ & $3000 - 10000$ & $0.1 - 1$ & $0.007 - 1.8$ \\ 
 m11d & 7100 & 0 & $2.6 \times 10^{11}$ & 6 & $0.001 - 0.01$ & $3000 - 10000$ & $0.1 - 10$ & $0.007 - 1.8$ \\ 
 m11e & 7100 & 0 & $1.4 \times 10^{11}$ & 5 & $0.001 - 0.01$ & $3000 - 10000$ & $0.1 - 10$ & $0.012 - 1.8$ \\ 
 m11f & $1.2 \times 10^4$ & 0 & $4.4 \times 10^{11}$ & 7 & $0.001 - 0.01$ & $3000 - 10000$ & $0.1 - 10$ & $0.007 - 1.8$ \\ 
 m11h & 7100 & 0 & $1.9 \times 10^{11}$ & 7 & $0.001 - 0.01$ & $3000 - 10000$ & $0.1 - 10$ & $0.007 - 1.8$ \\ 
 \hline
 \textbf{m11i} & 7100 & 0 & $6.7 \times 10^{10}$ & 113 & $0 - 0.1$ & $0 - 42500$ & $0 - 100$ & $0.003 - 18$ \\ 
 \hline
 m11q & 7100 & 0 & $1.5 \times 10^{11}$ & 27 & $0.0001 - 0.01$ & $1000 - 30000$ & $0.01 - 1$ & $0.007 - 9$ \\ 
 m10q & 250 & 0 & $9.8 \times 10^{9}$ & 9 & $0.001 - 0.01$ & $3000 - 10000$ & $0.1 - 1$ & $0.007 - 1.8$ \\ 
 \hline\hline
\end{tabular}
\caption{Parameters of simulation suite.  Columns include: \textit{(i) Halo name.}  The first 3 characters of each name roughly indicate halo mass.  The four most massive halos (listed first in this table) have been referred to in other papers simply as A1/2/4/8 (as part of the MassiveFIRE suite).  We append `m13' to the names in this paper for consistency with the other halos.  For all other halos, names are consistent with those in other FIRE papers.  \textit{(ii) Resolution} in units of \Msun, indicating the mass of baryonic resolution elements in each simulation.  \textit{(iii) $z_{\rm final}$}, the redshift at which each run halts.  \textit{(iv) $M_{\rm halo}$ at $z_{\rm final}$} in units of \Msun, the halo mass at the final snapshot, measured as $M_{200c}$.  \textit{(v) N$_{\rm runs}$}, the number of runs performed for each halo with different variations of SMBH physics.  \textit{(vi) $\eta_{\rm CR}$}, the range of values of cosmic ray feedback efficiency covered by the variations for that halo. \textit{(vii) $v_{\rm wind}$}, the range of values of wind velocity in km/s covered by the variations for that halo.  \textit{(viii) $\eta_{\rm RP}$}, the range of values of radiative feedback efficiency covered by the variations for that halo.  \textit{(ix) $\bar{\eta}_{\rm acc}$}, the range of values of effective accretion efficiency covered by the variations for that halo.}  
\label{tab:sims}
\end{center}
\end{table*}

\subsection{Black hole accretion models}
\label{ssec:accretion}

We implement and test several models for ``live" black hole accretion which describe the flow of gas into the black hole system from the surrounding medium in an accretion kernel of $\sim256$ gas resolution elements.  The black hole ``particle" represents both the black hole itself as well as an accretion disk with a separate reservoir of mass.  Mass flows from the surrounding gas into the accretion disk at a rate $\dot{M}_{\rm acc}$ determined by an accretion efficiency function $\eta_{\rm acc}$.  From the disk, it subsequently accretes onto the black hole or is expelled in a form of feedback (described in Section \ref{ssec:feedback}).  The accretion rate from the disk onto the black hole itself ($\dot{M}_{\rm BH}$) is motivated by the analytic \citet{Shakura1973} $\alpha$-disk model such that $\dot{M}_{\rm BH} = {M}_{\rm \alpha disk}/t_{\rm dep}$ where the depletion time $t_{\rm dep} = {\rm 42~Myr}(1+M_{\rm BH}/M_{\rm \alpha disk})^{0.4}$.  In practice, we find that the behavior of the model is not very sensitive to modest variations in either normalization or power-law index of $t_{\rm dep}$.  We then assume that a fraction $\epsilon_R = 0.1$ of this mass-energy is lost to radiation (see Section~\ref{ssec:feedback}).  In this formulation, the accretion rate becomes comparable to the Eddington limit when $M_{\rm BH} \sim M_{\rm \alpha disk}$ and is sub-Eddington for lower-mass disks.  We do not impose any additional Eddington limit on the accretion rate, but do not allow the accretion disk mass to exceed 10 times the BH mass.

In a preliminary set of simulation runs (not shown here in detail), we tested a wide variety of accretion models, including pure \citet{Bondi1944} accretion, torque-driven accretion, accretion only from cold bound gas, accretion with a fixed efficiency per free-fall time, high feedback efficiency at low accretion rates, etc., as well as numerical variations on these models.  We also systematically varied the scaling of $t_{\rm dep}$ and limit to $M_{\rm \alpha disk}$. We found that several models produced clearly unphysical or problematic behavior and would not be promising candidates for deeper study.  As one example, we found (as has also been shown in other numerical studies, see e.g. \citet{Angles-Alcazar2013}) that the unmodified Bondi-Hoyle accretion model is highly dependent on choice of BH seed mass: low-mass seeds accrete at too-low rates and never reach maturity, while high-mass seeds accrete too quickly and easily become over-massive.  From this initial survey, we identified four accretion models with physically-reasonable behavior which we employed in the larger suite of simulations discussed in this paper. Reassuringly, as we note below, this set includes models motivated by higher-resolution simulations and observations of accretion on sub-pc scales around AGN.

For each accretion model, the rate at which gas flows into the BH accretion disk can be expressed in the form
\begin{align} 
\dot{M}_{\rm acc} &= \eta_{\rm acc}\,M_{\rm gas}\,\Omega
\end{align}
where $M_{\rm gas}$ is the gas mass within the BH accretion kernel and $\Omega = \sqrt{G\,M_{\rm tot}/R^{3}}$ is the dynamical frequency at the force-softening radius calculated from the total mass enclosed $M_{\rm tot}$, while the accretion efficiency $\eta_{\rm acc}$ is a function which varies from model to model.  

To compare accretion efficiencies across different accretion models, we define an effective accretion efficiency
\begin{align} 
\bar{\eta}_{\rm acc} &= \left\langle \frac{\dot{M}_{\rm acc}}{M_{\rm gas}\,\Omega} \right\rangle
\end{align}
which is a single value that describes the average accretion efficiency over the course of each simulation.  We measure $\bar{\eta}_{\rm acc}$ from the simulations after they were run by comparing $\dot{M}_{\rm acc}(t)$ to $M_{\rm gas}(t)\,\Omega(t)$, where $M_{\rm gas}$ and $\Omega$ were measured in a 100-pc sphere around the BH at each snapshot.  We then identify the factor which will most closely bring these two timeseries into alignment for a given model.  The resulting values of $\bar{\eta}_{\rm acc}$ we report here are therefore comparable to one another within the scope of our analysis.  We caution, however, that because they have been renormalized they are not equivalent to the input values, and therefore should not be directly used in or compared to other studies.

The accretion models explored in more detail in this paper are as follows:

\begin{itemize}
    \item{\bf Fixed efficiency per free-fall time:} In this model, we assume gas accretes into the disk at a fixed rate per free-fall time at the force-softening radius of the BH particle,
    \begin{align} 
    {\eta}_{\rm acc} &= C \
    \end{align}
    or $\dot{M}_{\rm acc} = C\,M_{\rm gas}\,\Omega$ where $C$ is a chosen constant which can be varied between simulations.  The gas density at the location of at the black hole is calculated by interpolation of the cells in the accretion kernel, and the quantities above are derived by assuming that the density is constant out to the force-softening radius.
    \item{\bf Torque-driven:} This model assumes accretion is regulated by ``gravitational torques" in the galaxy arising from asymmetries in the gravitational potential, interactions between the collisionless (stars and dark matter) particles and the gas, and gaseous self-interaction (e.g., shocks and dissipation).  This scenario was studied in detail on sub-kpc scales by \citet{Hopkins2010, Hopkins2011}, and has been validated in simulations with a multi-phase ISM and stellar feedback with resolution down to $\sim 0.01$\,pc by  \citet{Hopkins2016, Angles-Alcazar2017b,  Angles-Alcazar2017, Angles-Alcazar2021}. From those studies, the typical accretion rate from $\gtrsim 10\,$pc scales can be parameterized by (see \citealt{Hopkins2011}): 
    \begin{align} 
    \label{eqn:H+Q}
    {\eta}_{\rm acc} &= C~\frac{(M_{\rm S}/M_{\rm d})^{1/6}}{1 + 3\,M_{\rm d,\,9}^{1/3}\,(M_{\rm gas}/M_{\rm d})} \
    \end{align}
    where $C$ is a chosen constant (which depends on e.g.\ the mass profile and star formation efficiency on unresolved scales), $M_{\rm S}=M_{\rm BH}+M_{\rm disk}$ is the total BH particle mass, $M_{\rm d}$ is the mass of ``disky" (angular-momentum-supported) material, and $M_{\rm d,\,9}\equiv M_{\rm d}/10^{9}$ \Msun.
    \item{\bf Isothermal sphere collapse:} This model assumes that the gas on scales within the accretion kernel is distributed in a spherically-symmetric, non-rotating isothermal sphere ($\rho \propto r^{-2}$) density profile, with a \citet{Bondi1944}-like accretion rate onto the BH from scales where the Bondi assumptions apply ($R \lesssim G\,M_{\rm BH}/(\sigma^{2}+v_{c}^{2})$), giving
    \begin{align} 
    \label{eqn:isothermal}
    \dot{M}_{\rm acc} &= C~\frac{4\pi G^2 M_{\rm tot}^2 \rho}{(\sigma^2 + v_c^2)^{1.5}} \\ 
    {\eta}_{\rm acc} &=  \frac{C}{(\sigma^2/v_c^2 + 1)^{1.5}} \
    \end{align}
    where $C$ is a chosen constant, $v_c = R\Omega$ is the circular velocity and $\sigma^2 = |{\bf v}_{\rm BH}-\bar{\bf v}_{\rm gas}|^2 + c_s^2$ is the effective velocity dispersion. Note that this model is a direct extension of the ``Shu solution" \citep{Shu1977} used to describe protostellar accretion in an isothermal medium (simply allowing for stars as well as gas and the point mass to contribute to the gravitational force).
    \item{\bf Shallow sphere collapse:} Here one makes the same assumptions as the isothermal/Shu solution model above, except to assume (ad-hoc) that the presence of a collisionless (stellar/dark matter) component leads the mass inside the kernel to be distributed in a shallower $\rho \propto r^{-1}$ density profile (as e.g.\ an NFW profile) outside the effective Bondi radius. This leads to exact solutions which do not have a simple closed analytic form, but can be well-approximated for all relevant limits here by:
    \begin{align} 
    \label{eqn:shallow}
    \eta_{\rm acc} &\approx \frac{C\,(M_{\rm BH}/M_{\rm tot})^{3/4}}{1 + (\sigma/v_{c})\,{\rm MIN}[(\sigma/v_{c})^{2},\, (M_{\rm tot}/M_{\rm BH})^{1/4}]}
    \end{align}
    \item{\bf Gravitational-acceleration dependence:} For each of the above models, we include a variant which accounts for the unresolved effects of stellar feedback on the scale of the accretion kernel by assuming that some fraction of material will be unbound by winds and therefore will not accrete.  The fraction of material which is actually available for accretion $f_{\rm acc}$ scales with the local gravitational acceleration $a_{\rm g,eff}$ as
    \begin{align}
    \label{eqn:fwind}
        f_{\rm acc} &= \frac{a_{\rm g,eff}}{a_{\rm g,crit} + a_{\rm g,eff}}
    \end{align}
    where $a_{\rm g,eff} = G M_{\rm enc} / R^2$ is the gravitational acceleration at the scale of the accretion kernel and $a_{\rm crit} \approx 10^{-7}\,{\rm cm\,s^{-2}}$ is the critical value below which a significant fraction of gas is expelled (as determined by the momentum-loading of stellar feedback, see \citet{Grudic2019} and \citet{Hopkins2022}, and corresponding to $M_{\rm enc}/\pi\,R^{2} \sim 3000\,M_{\odot}\,{\rm pc}^{-2}$).  In model variants which employ this $a_g$-dependence, the accretion efficiency is modified by the fraction of mass available for accretion,
    \begin{align}
        \eta_{\rm acc} \rightarrow f_{\rm acc}~\eta_{\rm acc}.
    \end{align}
    These models produce an effect similar to that described by \citet{Chen2020}, where galaxies with relatively low central surface densities have suppressed black hole growth and feedback.

\end{itemize}

\begin{figure*}
  \centering
  \includegraphics[width=2.1\columnwidth]{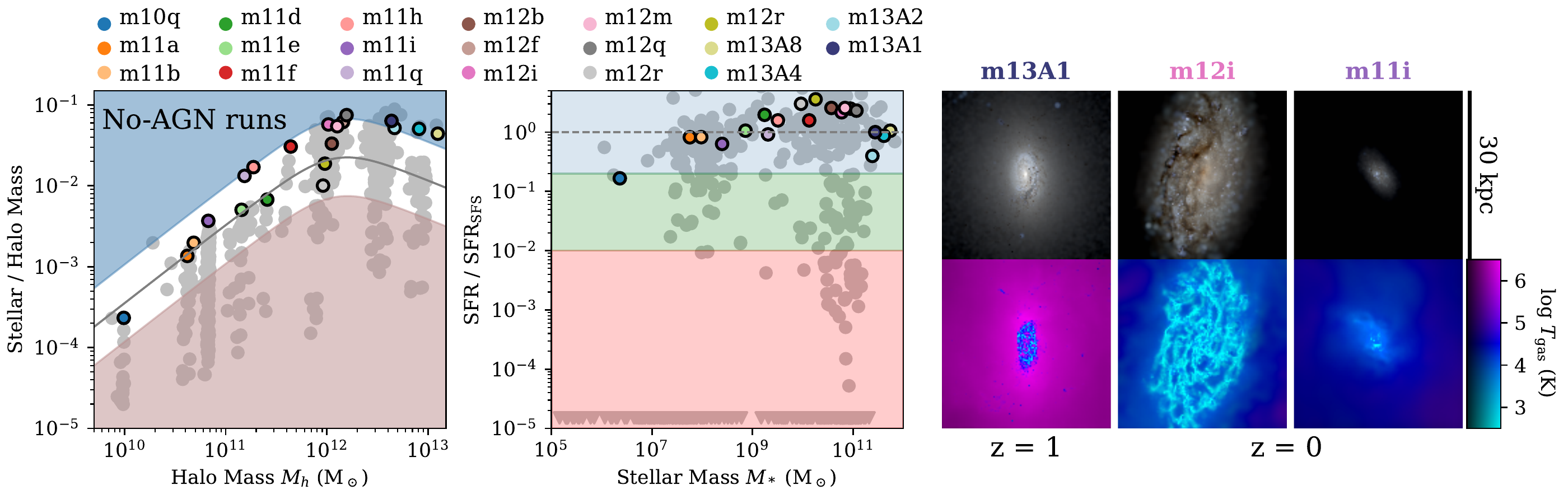}
  \caption{The stellar mass - halo mass (SMHM) relation (left) and star formation rate relative to the star-forming sequence (center) at the final snapshot of a comparable set of simulations which \textit{do not} include any SMBH physics.  These runs approximately follow known scaling relations up to halo masses of $\sim 10^{12}$ \Msun, but halos with masses above this value universally fail to quench and overshoot the SMHM relation.  (Light grey points refer to the simulations in our AGN suite shown in Figure \ref{fig:allsims}, and reference lines in grey are as defined as in the same Figure.)  Images of the gas and stars at the final snapshot of simulations m13A1, m12i, and m11i are shown on the right.  Galaxies at all masses have cold gaseous disks and ongoing star formation.}
  \label{fig:controlsims}
\end{figure*}

\subsection{Black hole feedback models}
\label{ssec:feedback}

Outflows are launched from the BH accretion disk system at a rate proportional to the accretion rate onto the black hole $\dot{M}_{\rm BH}$. We describe the different feedback physics here.

\subsubsection{Feedback Channels}

We implement three different channels for feedback, whose strengths can be varied independently:

\begin{itemize}
    \item{\bf Radiative Feedback:} We assume that a fraction $\epsilon_r = 0.1$ of the mass-energy accreted onto the black hole is converted into radiation, and follow multi-band radiation transport (with metallicity-dependent opacities) using the LEBRON method \citep{Hopkins2020}.  This is identical to how we treat radiative stellar feedback but adopts the empirical ``intrinsic'' (un-reddened) QSO template spectrum to calculate photo-ionization, photo-electric, and Compton heating/cooling rates \citep{Hopkins2016}.  The same radiation transport predicts the radiation pressure forces, i.e.\ momentum flux $\dot{p} = L_{\rm abs}/c$ where $L_{\rm abs}$ is the absorbed photon luminosity in a given gas element.  Our default treatment only includes the {\em explicitly resolved} radiation pressure, but we can also multiply the photon momentum by a factor $\eta_{\rm RP}$ to account for sub-structure below the resolution limit which may absorb radiation, such that
    \begin{align}
    \label{eqn:rp}
        \dot{p}_{\rm rad} &= \eta_{\rm RP}\,L_{\rm abs}\,/\,c
    \end{align}
    where $\eta_{\rm RP}$ is a parameter which adjusts the strength of radiative feedback.  Most of the simulations in our suite use values of $\eta_{\rm RP}$ between 0.1 and 10 (with a ``fiducial" value of 1) although we test values down to 0 and up to 100.
    
    \item{\bf Mechanical Feedback:} In addition to the matter which accretes onto the black hole itself, some matter will be launched away from the accretion disk in the form of winds, jets, or other outflows.  The outflows are physically generated at scales which are unresolved in our simulations, so we cannot model them directly and instead must describe their emergent properties at scales well outside the accretion disk system but before interaction with the galaxy.  We therefore parameterize the winds in a very general way to have mass outflow rates proportional to the black hole accretion rate $\dot{M}_{\rm out} = \eta_{\rm out}\,\dot{M}_{\rm BH}$ with $\eta_{\rm out} = 1$ and wind velocity $v_{\rm wind}$ at launch, such that the kinetic energy generated in the winds is
    \begin{align}
        \dot{E}_{\rm mech} = \dot{M}_{\rm BH}\,v_{\rm wind}^2\,/\,2
    \end{align}
    where $v_{\rm wind}$ is a parameter which adjusts the strength of mechanical feedback.  Most of the simulations in our suite use values of $v_{\rm wind}$ between 3,000 km/s and 10,000 km/s (with a ``fiducial" value of 10,000 km/s), although we test values down to 0 km/s and up to 42,500 km/s.  The wind velocity can also be expressed in terms of a mechanical feedback efficiency $\eta_{\rm mech} = (v_{\rm wind}/c)^2/2$ such that $\dot{E}_{\rm mech} = \eta_{\rm mech}\,\dot{M}_{\rm BH}\,c^{2}$.
    
    \item{\bf Cosmic Ray Feedback:} AGN are also known to accelerate massive particles to relativistic speeds, which then behave differently from the non-relativistic outflows.  Alongside the mechanical winds, we deposit cosmic rays (CRs) as a parallel ultra-relativistic fluid, similar to their injection during supernova events, and account for their diffusion, streaming, catastrophic and coulomb/ionization losses, adiabatic work and pressure-coupling terms between CRs and gas (see \citet{Chan2019} for details of the implementation). Once injected, all CRs are propagated following the identical equations (regardless of their source).  Cosmic rays are injected at a rate
    \begin{align}
    \label{eqn:crs}
    \dot{E}_{\rm CR} = \eta_{\rm CR}\,\dot{M}_{\rm BH}\,c^{2}. 
    \end{align}
    where the efficiency $\eta_{\rm CR}$ is a parameter which adjusts the strength of cosmic ray feedback.  Most of the simulations in our suite use values of $\eta_{\rm CR}$ between 0.001 and 0.01 (with a ``fiducial" value of 0.01), although we test values down to 0 and up to 0.1. 
\end{itemize}

\subsubsection{``Total'' Feedback Efficiency and ``Responsiveness''}

To parameterize the combined strength of AGN feedback in a given simulation, we define a total feedback efficiency
\begin{align}
    \label{eqn:etafb}
    \eta_{\rm fb} = \eta_{\rm CR,ref} \left( \frac{\eta_{\rm CR}}{\eta_{\rm CR,ref}} + \frac{\eta_{\rm mech}}{\eta_{\rm mech,ref}} + \frac{\eta_{\rm RP}}{\eta_{\rm RP,ref}} \right)
\end{align}
where the reference values $\eta_{\rm CR,ref} = 0.001$, $\eta_{\rm mech,ref} = 0.005$ (or $v_{\rm wind,ref}$ = 30,000 km/s), and $\eta_{\rm RP,ref} = 10$ have been chosen based on the approximate point at which each channel (in isolation) starts to suppress star formation in our massive galaxy simulations given reasonable black hole growth (some examples appear later in Figure~\ref{fig:greatsets}).  From these coefficients, it is apparent that we find cosmic ray feedback to be a few times more ``efficient" than mechanical feedback, in the sense that it requires less energy to effect a change in the galaxy-level star formation rate.  Because it is often the case within our suite that $\frac{\eta_{\rm CR}}{\eta_{\rm CR,ref}}$ is larger than either $\frac{\eta_{\rm mech}}{\eta_{\rm mech,ref}}$ or $\frac{\eta_{\rm RP}}{\eta_{\rm RP,ref}}$, we normalize Eqn.~\ref{eqn:etafb} such that $\eta_{\rm fb} \approx \eta_{\rm CR}$ when the cosmic ray term dominates.

For each simulation in the suite, we also define a ``responsiveness" parameter
\begin{align}
    \label{eqn:responsiveness}
    \eta_{\rm R} = \bar{\eta}_{\rm acc}\eta_{\rm fb}
\end{align}
which combines the accretion and feedback efficiencies to indicate how responsive the AGN model is to changes in the local environment.  We discuss the significance of this parameter further in Section~\ref{sec:sfreg}.

\subsubsection{Feedback Coupling: Numerics and Geometry}

In addition to varying the strength of feedback in these three channels, we vary the numerical implementation employed to inject the feedback and physical assumptions about the coupling geometry.  We test the following four schemes:
\begin{itemize}
    \item{\bf Push:} Feedback is injected isotropically into the gas cells within the interaction kernel, weighted by the solid angle each cell subtends from the perspective of the BH particle (similar to the feedback scheme for stars, see \citet{Hopkins2016} for implementation details).
    \item{\bf PushV:} Feedback is injected isotropically into the gas cells within the interaction kernel, weighted by the volume of each cell.
    \item{\bf Spawn:} New resolution elements (with $10^{-2}$ times the mass of other particles) are created in the vicinity of the BH particle to represent the outflow and stream outwards isotropically.  Resolution elements are created in pairs moving in opposite directions to conserve momentum.  They obey the normal equations of motion until they encounter a ``normal" cell within their hydrodynamic kernel with a similar velocity vector, at which point they may merge and combine their energy and momentum.  (See \citet{Torrey2020} for implementation details.)
    \item{\bf Jet:} Numerically identical to ``spawn," but all particles are injected along the $\pm z$-axis (as defined by the combined angular momentum vector of accreted material), so are effectively infinitely collimated at launch.  (See \citet{Su2021} for implementation details.)
\end{itemize}

\subsection{Derived quantities}

For each snapshot output, we calculated several bulk properties of the galaxy such as its stellar mass, velocity dispersion, and star formation rate.  Basic snapshot-reading functions and particle manipulation and the identification of the baryonic center of the halo were carried out using the GizmoAnalysis package \citep{Wetzel2020}.  We measure the following quantities with the coordinate system rotated such that the $z$-axis is aligned with the angular momentum vector of the stars within 10 kpc of the baryonic center: 
\begin{itemize}
    \item Stellar mass $M_*$: the total mass of star particles within 50 kpc of the baryonic center.
    \item Star formation rate SFR: the total mass of stars within 50 kpc which were formed within the past 300 Myr, divided by 300 Myr.
    \item Effective radius $R_e$: the two-dimensional radius enclosing stellar mass $M_*/2$ when the galaxy is viewed face-on (i.e., projected along the $z$-axis).
    \item Velocity dispersion $\sigma$: the standard deviation of the $z$-velocity of stars within $R_e$.
    \item Halo mass $M_h$: the total mass $M_{200c}$ inside the radius $R_{200c}$ which encloses a sphere whose average density is 200 times the critical density.
\end{itemize} 

\begin{figure*}
  \centering
  \includegraphics[width=2\columnwidth]{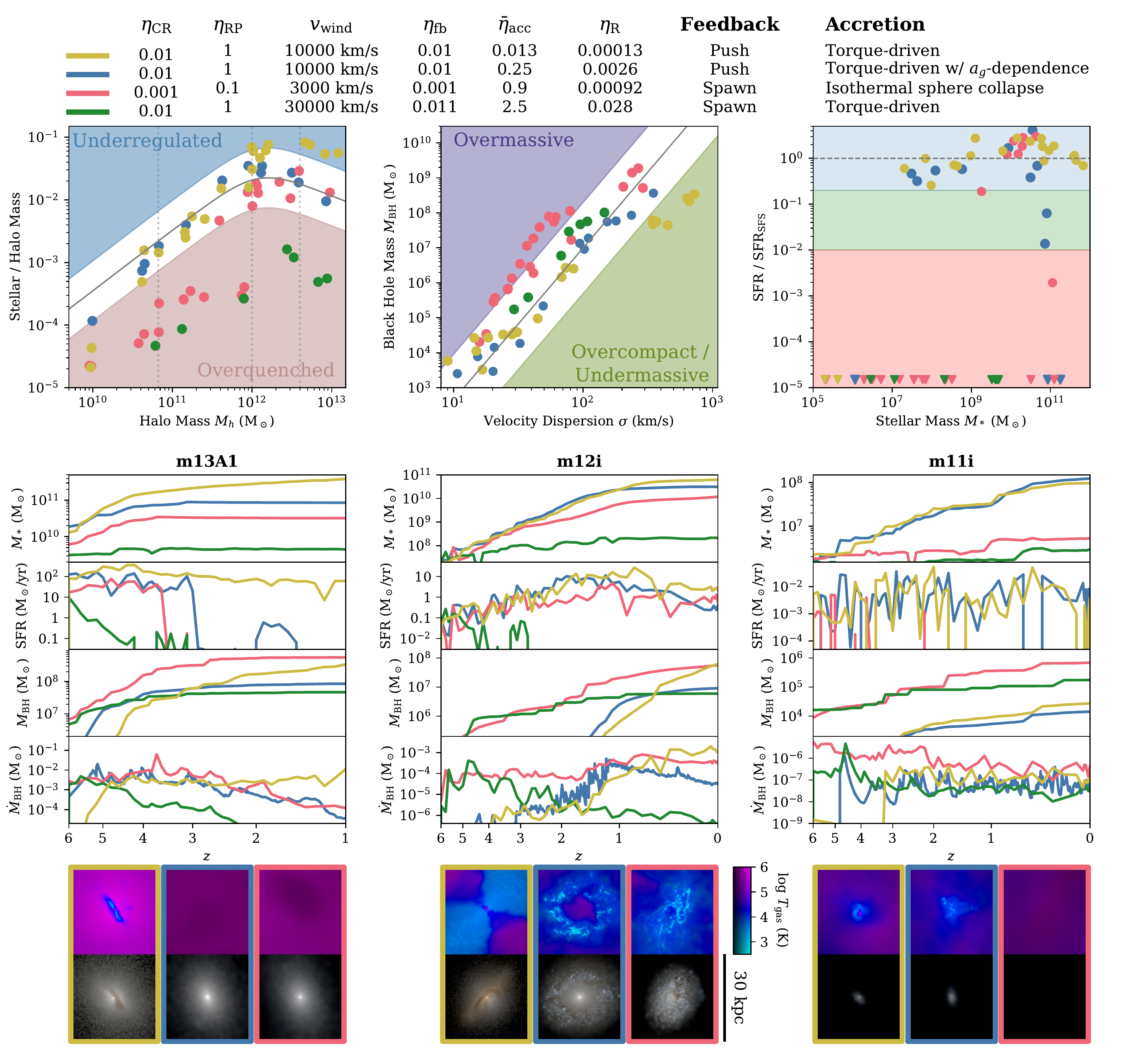}
  \caption{Four sets of simulations utilizing different AGN models and parameterizations, demonstrating the variety of quenching-related outcomes across halo mass.  Line/symbol color indicates which model (listed at the top of the figure) was used in that run, including a model which fails to quench (gold), quenches massive galaxies (blue), quenches massive and dwarf galaxies (pink), and quenches all galaxies (green).  {\it Top row:} Scaling relations (stellar mass - halo mass relation, $M-\sigma$ relation, and star formation rate relative to the star forming sequence) as defined in Figure~\ref{fig:allsims}, at the final snapshot of each simulation.  Vertical dotted lines in the top left panel mark the final halo mass for m13A1, m12i, and m11i.  {\it Middle rows:} Time evolution of stellar mass, star formation rate, black hole mass, and black hole accretion rate for each model variation in haloes m13A1, m12i, and m11i.  {\it Bottom:}  Images of the gas density and temperature (first row) and stellar light (second row) at the final snapshot of haloes m13A1, m12i, and m11i with the first three model variations.}
  \label{fig:varietysets}
\end{figure*}

\section{Outcomes of the simulation suite}
\label{sec:results}

The properties of the final snapshot for each simulation in the suite ($z=0$ for m12X and m11X, and $z=1$ for m13X), can be seen in Figure~\ref{fig:allsims}.  Points are colored by halo identity, with all points of the same color having the same initial conditions.  The leftmost panel shows the stellar mass - halo mass (SMHM) relation, indicating the level of star formation regulation in each halo.  For comparison, the thin grey line shows the observationally-calibrated $z=0$ SMHM relation from \citet{Behroozi2019}.  (We note that the SMHM relation is not observed to evolve substantially with redshift, so the $z=0$ relation is relevant for points both at $z=0$ and $z=1$.)  Points well above this line (in the blue-shaded region) have higher star formation efficiencies than expected, while points well below it (in the red-shaded region) have formed fewer stars than expected.  While the majority of the simulations have final stellar masses that are in line with expectations, across all halo masses there are also many simulations which over-suppress star formation.  Conversely, several simulations at the high-mass end fail to sufficiently regulate their star formation and overshoot the relation.  The remainder of this paper will be dedicated to elucidating how the choice of SMBH physics affects where simulated galaxies land on this diagram, and whether and how that changes with halo mass.

Another important scaling relation to consider is $M$-$\sigma$, which relates a property of the SMBH ($M_{\rm BH}$) to a property of the galaxy ($\sigma$, stellar velocity dispersion).  This relationship is well-established in the local Universe (see e.g. \citet{Kormendy2013} for a review) and points to a physical relationship between galaxies and SMBHs.  The center panel of Figure~\ref{fig:allsims} shows $M$-$\sigma$ for all simulations in our suite.  The grey line shows a fit to observational data from \citet{Greene2020}.  Most of the simulations lie within a factor of 10 of this relation, but there are also many runs with overmassive black holes which lie well above it.  At the high-mass end, some simulations lie below the relation, indicating either that the black hole is undermassive or that the galaxy is overly compact ($\sigma$ is too high).

Finally, we consider the amount of active star formation in the final 300 Myr of each simulation.  The rightmost panel shows the average star formation rate (SFR) during this period of time, normalized against the median value for galaxies at that mass and redshift on the star-forming sequence from \citet{Whitaker2014} at $z > 0.5$ and \citet{Salim2007} at $z=0$ (shifted according to the average corrections to stellar mass and SFR suggested by \citet{Leja2019}).  Many simulations lie near the median value and would be considered actively star-forming, others have suppressed star formation and lie below it, while many more have SFRs several orders of magnitude below the star-forming sequence value.  In this paper, we use the word ``quenched" loosely to refer to a galaxy's position on this diagram: we consider a galaxy quenched when it lies in the red region, at least two orders of magnitude below the star-forming sequence.  Note that this is a temporary definition that refers to the state at the final snapshot of each simulation, rather than a statement about long-term behavior.  (And indeed, because the massive halos were run only to $z=1$, we are not able to test whether quenching is long-lasting down to $z=0$.)
In contrast, we use the word ``regulation" to refer to a galaxy's position with respect to the SMHM relation.  Most massive galaxies which are ``well-regulated" are also ``quenched" -- but the reverse is not necessarily the case.

For comparison, consider the simulations shown in Figure~\ref{fig:controlsims} which have the same initial conditions and resolution as our suite, but do not include any SMBH physics.  Up to a halo mass of $M_h \approx 10^{12}$ \Msun~the simulations follow many of the known galaxy scaling relations, including the SMHM relation as shown in the leftmost panel.  At higher masses, however, the simulations without SMBH physics fail to adequately regulate star formation and overshoot the SMHM relation, as discussed in detail by \citet{Su2019}.  They continue steadily forming stars, as can be seen in the central panel, and do not quench.  Moreover, the simulated massive galaxies build up very high central stellar densities which place them well below the observed size-mass relation at $z < 2$ \citep{Parsotan2020}. Other simulation studies have found similar central stellar overdensities when AGN feedback is not included, see e.g. \citet{Choi2018}.

The images on the right of Figure~\ref{fig:controlsims} show the three halos that were run with the most physics variations in our simulation suite, one per decade of mass: m13A1 (a very massive, early-forming galaxy), m12i (a Milky-Way-like galaxy), and m11i (a dwarf galaxy).  The top row of images show stellar light, while the bottom row show the underlying gas temperature and density with cold gas in cyan and hot gas in magenta.  The brightness of the color in the bottom panels indicates gas surface density.  Each galaxy is still actively star-forming at the conclusion of the no-AGN simulation regardless of mass, with m13A1 forming a hot halo with a nuclear disk surrounding a dense stellar core and m12i forming an extended spiral disk.

Within the AGN physics simulation suite, there are a variety of behaviors (as was shown in Figure~\ref{fig:allsims}) -  there are some runs with SMBHs which still fail to quench massive galaxies, while other models over-quench galaxies across halo mass, and some models are able to appropriately regulate star formation (but not necessarily at all masses).  Figure~\ref{fig:varietysets} shows four specific SMBH models which illustrate this variety of behaviors within the simulation suite (selected for display because they were run on a relatively large number of halos).  The top row of panels replicate the SMHM relation, $M$-$\sigma$ relation, and star formation rates from Figure~\ref{fig:allsims} but select only the subset of simulations belonging to one of the four models denoted at the top of the figure.  Points with the same color use the same SMBH accretion model, feedback model, and parameterization, but have different initial conditions.  The middle set of panels show the time evolution of halos m13A1, m12i, and m11i for each of the four models, and the images at the bottom of the figure depict the gas and stars at the final snapshot of those runs (comparable to the images in Figure~\ref{fig:controlsims}).

The first model, shown in gold, uses a torque-driven accretion model and a ``push" feedback scheme.  Feedback is moderately powerful and spread across all channels, but the effective accretion efficiency is low.  As a result, the BHs in the massive galaxies are too small for their feedback to be capable of affecting star formation.  Despite clearly visible outflows in the image for m12i, this model is indistinguishable from the no-AGN scenario in terms of the stellar mass growth of the galaxies.

The second model, shown in blue, is similar to the first but has a much higher effective accretion efficiency and includes the $a_g$-dependent prefactor which suppresses SMBH accretion at low surface densities.  Because the BH can accrete much more efficiently at high surface densities, the feedback from that accretion is more effective at regulating star formation in m13A1, while the $a_g$-dependence prevents overfueling in m11i.  This model is one of the few which produces BHs and galaxies which are in agreement with observational constraints across the full range of halo mass.  (We discuss some of the elements which lead to its success, and describe other successful models, in the later Sections of this paper.)

The third model, shown in pink, uses an isothermal sphere collapse accretion prescription and a particle-spawning feedback scheme.  The feedback is weaker in all three channels than in the previous models.  In this case, weakening the feedback efficiency has the counter-intuitive effect of {\it increasing} the amount of quenching, particularly in dwarf galaxies: the weakened feedback allows the SMBH to accrete more effectively and become overmassive, thus over-producing feedback which overpowers the star formation.  This model is also able to affect star formation in the more massive systems, quenching m13A1 and slightly suppressing star formation in m12i.

The final model, shown in green, uses a torque-driven accretion model, a particle-spawning feedback scheme, feedback efficiencies similar to the first two examples with somewhat stronger mechanical feedback, and very high effective accretion efficiency.  This model overquenches all galaxies across halo mass, thoroughly extinguishing star formation very early in all cases.  In this case, the overquenching is due not to over-massive black holes (as in the previous example), but rather to the extreme responsiveness of the model to accretion events.  The relatively strong feedback coupled with extremely high accretion efficiency means that the SMBH responds immediately and powerfully to fluctuations in its environment, driving impulsive feedback whenever an accretion event occurs.

In the following sections, we explore the general trends within the simulation suite -- the importance of regulating BH growth, the necessary balance of accretion and feedback efficiencies, and the importance of a model's responsiveness to its environment -- which are exemplified by these four models.

\section{Regulation of BH growth}
\label{sec:bhreg}

\begin{figure}
  \centering
  \includegraphics[width=\columnwidth]{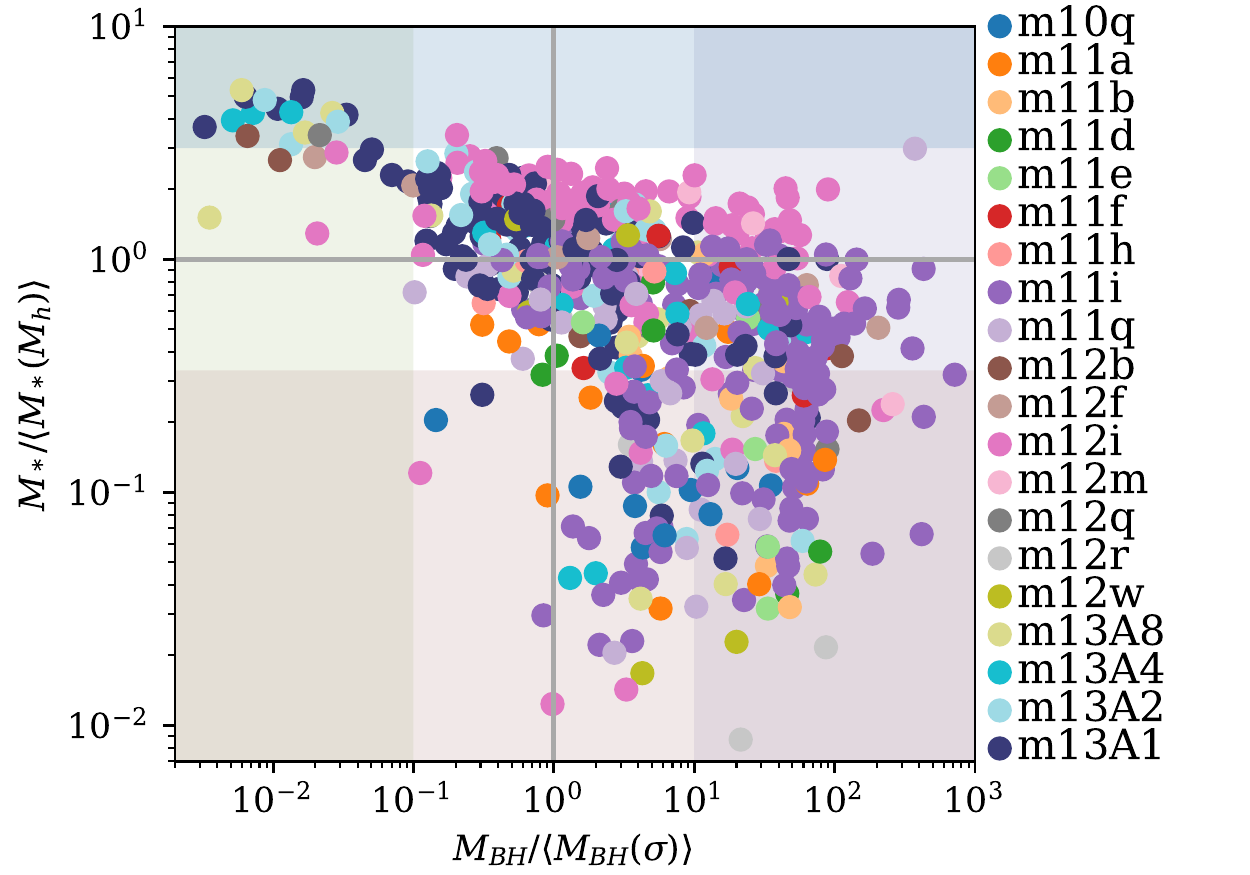}
  \caption{Regulation of BH growth vs. regulation of star formation in our simulated galaxies, as represented by the distance from the $M-\sigma$ relation (x-axis) and stellar mass - halo mass relation (y-axis).  Each point represents the state of the system at the final snapshot of a single simulation, with the color indicating the halo identity.  (Multiple points with the same color have different models and parameterizations of BH physics.)  Many points are clustered around $x=y=1$ where both the BH and stellar mass are well-regulated, but there are also several simulations with overmassive black holes and over-quenched galaxies (lower right quadrant) or undermassive black holes and under-regulated galaxies (upper left quadrant), indicating that self-regulation of BH growth is relevant to the regulation of star formation in galaxies.} 
  \label{fig:bhmstar}
\end{figure}

\begin{figure*}
  \centering
  \includegraphics[width=2\columnwidth]{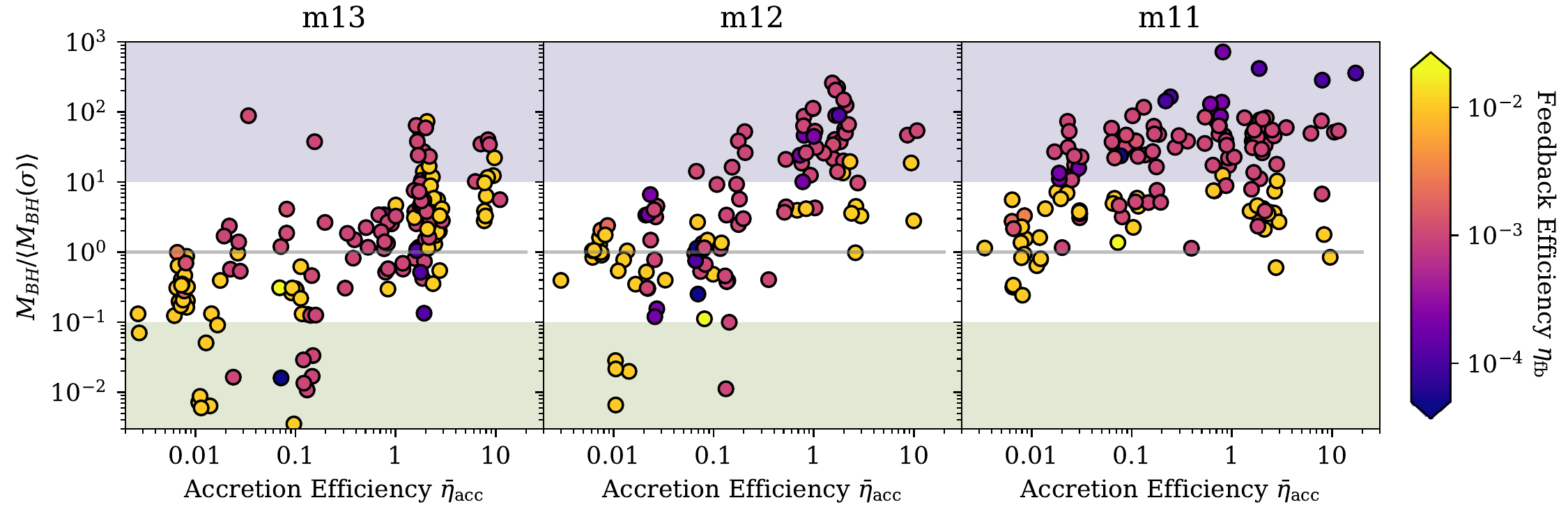}
  \caption{The regulation of black hole growth (y-axis, vertical distance from the $M-\sigma$ relation) depends on the both the accretion (x-axis) and feedback (color) efficiencies, and this dependence varies by halo mass.  A small amount of jitter has been introduced in the x-direction to separate points with identical effective accretion efficiencies.  In halos above $10^{12}$ \Msun, the final black hole mass scales predominantly with the effective accretion efficiency.  At lower masses, the final BH mass is determined almost entirely by the feedback efficiency, with little impact from the effective accretion efficiency.  The large scatter evident in this plot is driven in part by the differences between the properties of different pairs of feedback and accretion models - they obey the same qualitative trends but differ in detail.}
  \label{fig:bhreg}
\end{figure*}

\begin{figure*}
  \centering
  \includegraphics[width=2\columnwidth]{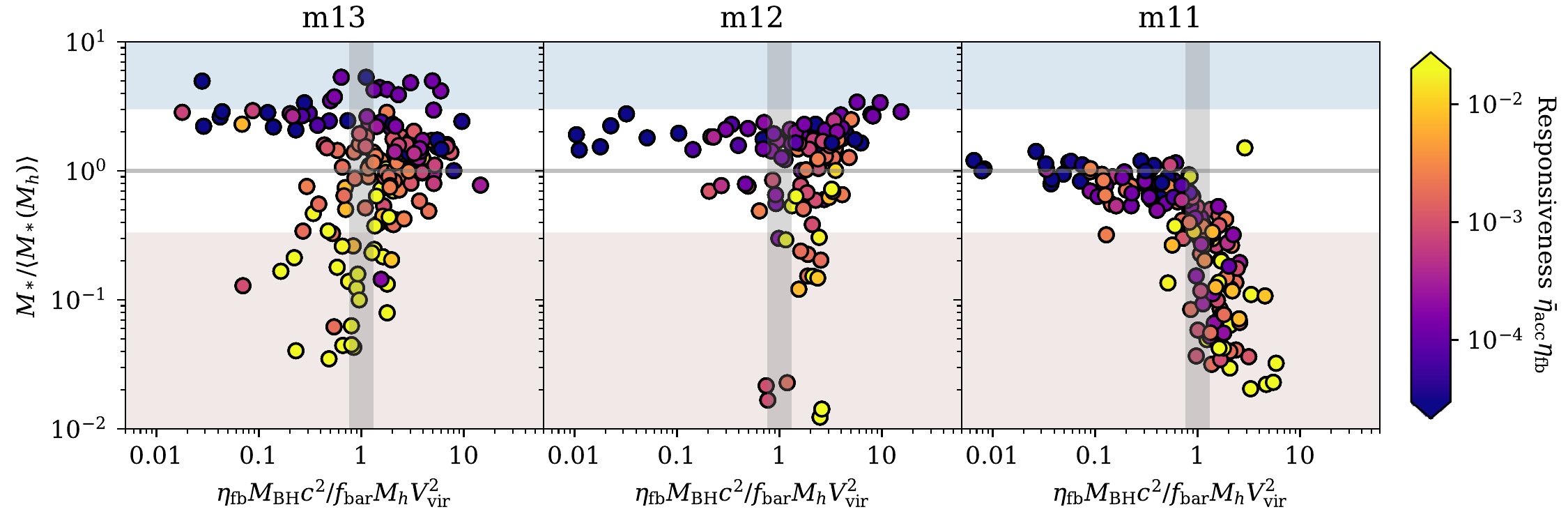}
  \caption{The regulation of star formation in galaxies (y-axis, vertical distance from the stellar mass - halo mass relation) depends both on the total amount of energy from AGN feedback (x-axis) and the manner in which that energy is injected (color).  The integrated energy from BH feedback is quantified on the x-axis as $\eta_{\rm fb} M_{\rm BH} c^2$ relative to the binding energy of the baryons in the halo, $f_{\rm bar} M_h V_{\rm vir}^2$.  In dwarf galaxies, star formation is over-suppressed once the energy from feedback exceeds the baryonic binding energy.  In more massive systems, whether star formation is suppressed depends also on {\it how} that energy was injected, as indicated by a vertical color gradient.  Color shows ``responsiveness" ($\bar{\eta}_{\rm acc} \eta_{\rm fb}$), a parameter indicating how quickly and forcefully the BH responds to the presence of gas available for accretion. In massive systems, the responsiveness of the model affects the level of quenching, and the most highly responsive models can overquench the galaxy even while injecting less energy overall.}
  \label{fig:sfreg}
\end{figure*}

The behavior of the third (pink) model in Figure~\ref{fig:varietysets} (where overmassive black holes lead to over-quenching of star formation in low-mass galaxies) suggests that the regulation of star formation within galaxies is related to the self-regulation of the growth of the black hole itself.  Figure~\ref{fig:bhmstar} demonstrates this general trend over all the simulations in our sample.  The final snapshot of each simulation is represented as a single point on this plot, with the x-axis value $M_{\rm BH}/\langle M_{\rm BH}(\sigma) \rangle$ indicating the regulation of BH growth by the vertical distance from the $M-\sigma$ relation, and the y-axis value $M_*/\langle M_*(M_h) \rangle$ indicating the regulation of star formation by the vertical distance from the SMHM relation.  

Many points are clustered near the crossing of the grey lines at $x=y=1$, where both the BH mass and stellar mass are well-regulated -- an encouraging (and not necessarily expected) result given the wide variety of models and parameter choices employed.  However, other regions of the plot are also populated.  In particular, the lower right quadrant is populated with many simulations resembling the low-mass galaxies in the third model of Figure~\ref{fig:varietysets}, which have overmassive black holes and undermassive galaxies.  The opposite (upper left) quadrant is also populated with simulations which have undermassive black holes and overmassive galaxies.  These are mostly high-mass systems where the BH feedback is not sufficiently powerful to suppress star formation.  Clearly, the ability of a given set of SMBH physics to regulate the growth of the BH itself is relevant to its ability to regulate star formation in galaxies - a point to which we will return in Section~\ref{sec:sfreg}.

How do our choices about SMBH physics affect BH growth?  The rate at which the BH can grow is a contest between the accretion and feedback efficiencies: high accretion efficiencies increase the ability of the BH to claim matter in its vicinity, but high feedback efficiencies can reduce the amount of matter available for accretion.  In Figure~\ref{fig:bhreg} we demonstrate how this balance between accretion and feedback manifests in our simulation sample as a function of halo mass.  We separate the sample into the m13X simulations (which reach halo masses of $4-12 \times 10^{12}$ \Msun~at $z=1$), m12X simulations (which reach halo masses of $\sim10^{12}$ \Msun~at $z=0$), and m11X simulations (which reach halo masses of $4-40 \times 10^{10}$ \Msun~at $z=0$), and show the final black hole mass normalized by its $M-\sigma$ value as a function of effective accretion efficiency.

In the higher-mass (m13 and m12) halos, there is a trend between these two quantities: higher effective accretion efficiencies produce higher-mass BHs, as one might expect.  There is significant scatter in this relationship due to differences in the detailed behavior of the various accretion and feedback implementations, as well as variations among halos with different initial conditions, but the overall trend holds despite these variations.  A third dimension also contributes to the scatter in the relation: overall feedback efficiency (Eqn.~\ref{eqn:etafb}), indicated by color.  

The effect of feedback on BH self-regulation is especially apparent in the lower-mass (m11) simulations, where the trend with accretion efficiency is much weaker, and the feedback efficiency becomes the dominant determinant of the final BH mass (visible as a vertical color gradient).  In these halos, models with stronger feedback are more effective at preventing black hole growth.  We also find that relatively strong feedback is \textit{required} to prevent the overgrowth of BHs in these dwarf galaxies (and subsequent overquenching of star formation) once the effective accretion efficiency exceeds a value of 0.03.  

This result may appear to be in tension with previous works from the FIRE collaboration indicating that BHs in low-mass galaxies are expected to be \textit{undermassive}, e.g. \citet{Angles-Alcazar2017, Catmabacak2022}, Byrne et al (in prep).  However, there are several salient differences between those studies and what we present here which limit direct comparison.  First (and perhaps most significant), here we use the $M-\sigma$ relation to define ``overmassive" black holes, while those studies use $M_{\rm BH}-M_*$.  We have examined how our results change if we use $M_{\rm BH}-M_*$ instead, and found that most of the overmassive BHs in dwarf galaxies (relative to $M-\sigma$) would {\it not} be considered overmassive relative to $M_{\rm BH}-M_*$.  Thus the two relations appear to make different predictions when extrapolated down to dwarf galaxies.  This difference between the low-mass predictions made by the two relations is consistent with observational results by \citet{Reines2015} and \citet{Baldassare2020} who measure $M_{\rm BH}-M_*$ and $M-\sigma$ respectively down to the dwarf galaxy regime and find that the BHs in dwarf galaxies lie below $M_{\rm BH}-M_*$ but along $M-\sigma$ when the relations are extrapolated to the low-mass end.  

There are also a number of differences between the simulations studied in this work and those in the previous works which may affect the long-term BH growth behaviors.  The simulations studied in previous works did not include any BH feedback which would self-regulate the BH growth.  Because of this, in order to reach similar BH masses they generally employed models with lower accretion efficiencies than we have used in this suite.  Additionally, they have primarily examined the high-redshift dwarf progenitors of present-day massive galaxies, while we refer to present-day dwarfs. Finally, there are some subtle but possibly significant differences in implementation between the simulations (e.g., the accretion disk reservoir and the centering method).  For all these reasons, we do not consider our results incompatible with what has been shown in previous work.

Thus we find that at high halo masses, the rate of BH growth is most dependent upon the chosen accretion efficiency, while at low halo masses it is most dependent upon the feedback efficiency.  We posit that this difference may be a result of the differing shapes of the central gravitational potentials in these systems: in dwarf galaxies with shallow central potentials, outflows can easily arrest the flow of gas into the vicinity of the black hole, while in more massive systems with deeper central potential wells and more well-formed disks, the outflows are more likely to be overcome or redirected such that inflowing gas is more able to reach the central region and be available for accretion. 

The balance point between accretion and feedback efficiencies that appropriately regulates BH growth therefore depends not only on those efficiencies, but on the halo mass (and in finer-grained detail, on the implementation of accretion and feedback).  For a given pair of accretion and feedback models, it is not necessarily the case that a ``solution" exists which satisfies this balance for all masses.  We find that models that incorporate an $a_g$-dependence for accretion (as described in Section~\ref{ssec:accretion}) are more likely to be able to achieve this balance because they naturally suppress BH overgrowth in dwarf galaxies without requiring the invocation of very strong feedback which could result in over-quenching of higher-mass galaxies.

\section{Regulation of star formation}
\label{sec:sfreg}

Clearly, achieving self-regulation of BH growth is an important first step to achieving regulation of star formation.  We have seen that BHs which are allowed to grow too large can overquench dwarf galaxies, while BHs which grow too sluggishly in massive galaxies will be unable to quench them.  To first order, the mass of the black hole and the feedback efficiency determine the cumulative amount of mass-energy that has been injected through feedback, i.e., $E_{\rm BHFB} \sim \eta_{\rm fb} M_{\rm BH} c^2$.  We expect that this quantity should therefore be an important factor in star formation regulation.

Figure~\ref{fig:sfreg} shows how the level of star formation regulation in our simulated galaxies (as quantified by vertical distance from the SMHM relation) changes as a function of $E_{\rm BHFB}$.  In the figure, we normalize this quantity by the binding energy of the baryons in the halo, $f_{\rm bar} M_h V_{\rm vir}^2$ (where $f_{\rm bar}$ is the universal baryon fraction and $V_{\rm vir}$ is the virial velocity of the halo) to be able to compare systems with slightly different halo masses.  As in Figure~\ref{fig:bhreg}, we separate the simulations by decade of halo mass into m13s, m12s, and m11s.  In the low-mass galaxies (the rightmost panel) there is a clear relationship between $E_{\rm BHFB}$ and $M_*/M_*(M_h)$: at low total feedback energies the stellar mass growth is barely affected, but when the energy released in AGN feedback exceeds the binding energy of the halo, it rapidly starts to overquench.  

For the more massive galaxies, however, the outcome is more complicated - the total amount of feedback energy is not sufficient to identify the systems whose star formation will be suppressed.  Rather, we find that {\it how and when} that feedback energy is released is also important.  The burstiness (or duty cycle) of feedback is not explicitly modeled in our implementation of SMBH physics.  However, we can consider a related property which is an intrinsic attribute of each model: its {\it responsiveness} to changes in the availability of gas for accretion.  If we consider a scenario where there is a sudden inflow of gas in the vicinity of the SMBH, the rapidity with which the SMBH can respond and the amount of mass it can accrete in a short timeframe are governed by the accretion efficiency $\bar{\eta}_{\rm acc}$.  The feedback energy per mass accreted is governed by the feedback efficiency $\eta_{\rm fb}$.  Therefore, the total amount of feedback energy from that accretion event would be proportional to $\bar{\eta}_{\rm acc}\eta_{\rm fb} \equiv \eta_{\rm R}$.  We term this quantity $\eta_{\rm R}$ ``responsiveness," as it indicates how quickly and powerfully the BH will respond to changes in its environment.  This quantity is related to, but is not the same as, burstiness: it does not guarantee that burst events {\it will} occur (these depend also on the circumstances of the given halo), but it does indicate how powerful a burst could be {\it if} it were to occur.  Responsiveness is an intrinsic attribute of each model, while burstiness describes a potentially emergent behavior.

We color the points in Figure~\ref{fig:sfreg} by the responsiveness $\bar{\eta}_{\rm acc}\eta_{\rm fb}$ of the SMBH model employed in each simulation.  In the most massive galaxies (left panel), a color gradient is visible in the vertical direction, indicating that at fixed total feedback energy, more responsive models are more effective at suppressing star formation.  Models with extremely high responsiveness (yellow points) can suppress BH growth along with star formation such that they end up with relatively low total feedback energy despite having overquenched the galaxy.  The fourth (red) model in Figure~\ref{fig:varietysets} is an example of such behavior.  Models with low responsiveness (purple points) can fail to regulate star formation in massive galaxies even with relatively high total feedback energies -- in these cases, the feedback was likely introduced in a more constant and gentle stream, and gradually carved out a path to escape the galaxy rather than violently disrupting it.  We find that responsiveness $\eta_{\rm R}$ is a better predictor of quenching in massive galaxies than either $\bar{\eta}_{\rm acc}$ or $\eta_{\rm fb}$ alone, although $\bar{\eta}_{\rm acc}$ alone is a stronger predictor than $\eta_{\rm fb}$ alone.  As in Figure~\ref{fig:bhreg}, the scatter in these diagrams stems from differences in the detailed behavior of each pair of accretion and feedback models.

The relationship between responsiveness and galaxy quenching can be further understood by roughly comparing the timescale of feedback to the dynamical time $t_{\rm dyn}$.  The amount of energy released in a feedback event occurring over a timescale $\Delta t$ is $E_{\rm fb} = \eta_{\rm fb} \dot{M}_{\rm BH} c^2 \Delta t$.  In our accretion models, $\dot{M}_{\rm BH} \approx \bar{\eta}_{\rm acc} M_{\rm gas} \Omega$, so $E_{\rm fb} = \eta_{\rm fb} (\bar{\eta}_{\rm acc} M_{\rm gas} \Omega) c^2 \Delta t = \eta_{\rm R} M_{\rm gas} c^2 (\Delta t/t_{\rm dyn})$ where $M_{\rm gas}$ and $\Omega = 1/t_{\rm dyn}$ are measured locally to the black hole such that $M_{\rm gas}$ is roughly comparable to $M_{\rm BH}$ during high-accretion events.  If that feedback event were to release enough energy to unbind the gas in the galaxy over a timescale $\Delta t_{\rm crit}$, then $E_{\rm fb} \sim M_{\rm galaxy} v_{\rm esc}^2$ such that $\Delta t_{\rm crit}/t_{\rm dyn} = (M_{\rm galaxy}/M_{\rm BH})(v_{\rm esc}/c)^2 / \eta_{\rm R} \sim (0.03/\eta_{\rm R}) (M_h/10^{12}~{\rm M}_\odot)^{2/3}$.  When $\Delta t_{\rm crit}/t_{\rm dyn} \gg 1$, the galaxy has many dynamical times to adjust to the injection of feedback energy, but when $\Delta t_{\rm crit}/t_{\rm dyn} \lesssim$ a few, the energy is released too quickly for the galaxy to respond.  Our dwarf galaxy simulations nearly all fall in the latter regime, so the main determinant of quenching is whether the feedback energy exceeds the binding energy of the halo or not.  For more massive galaxies, however, quenching (or more generally, star formation regulation) depends not only on exceeding this threshold but on the value of $\eta_{\rm R}$.

\begin{figure*}
  \centering
  \includegraphics[width=2\columnwidth]{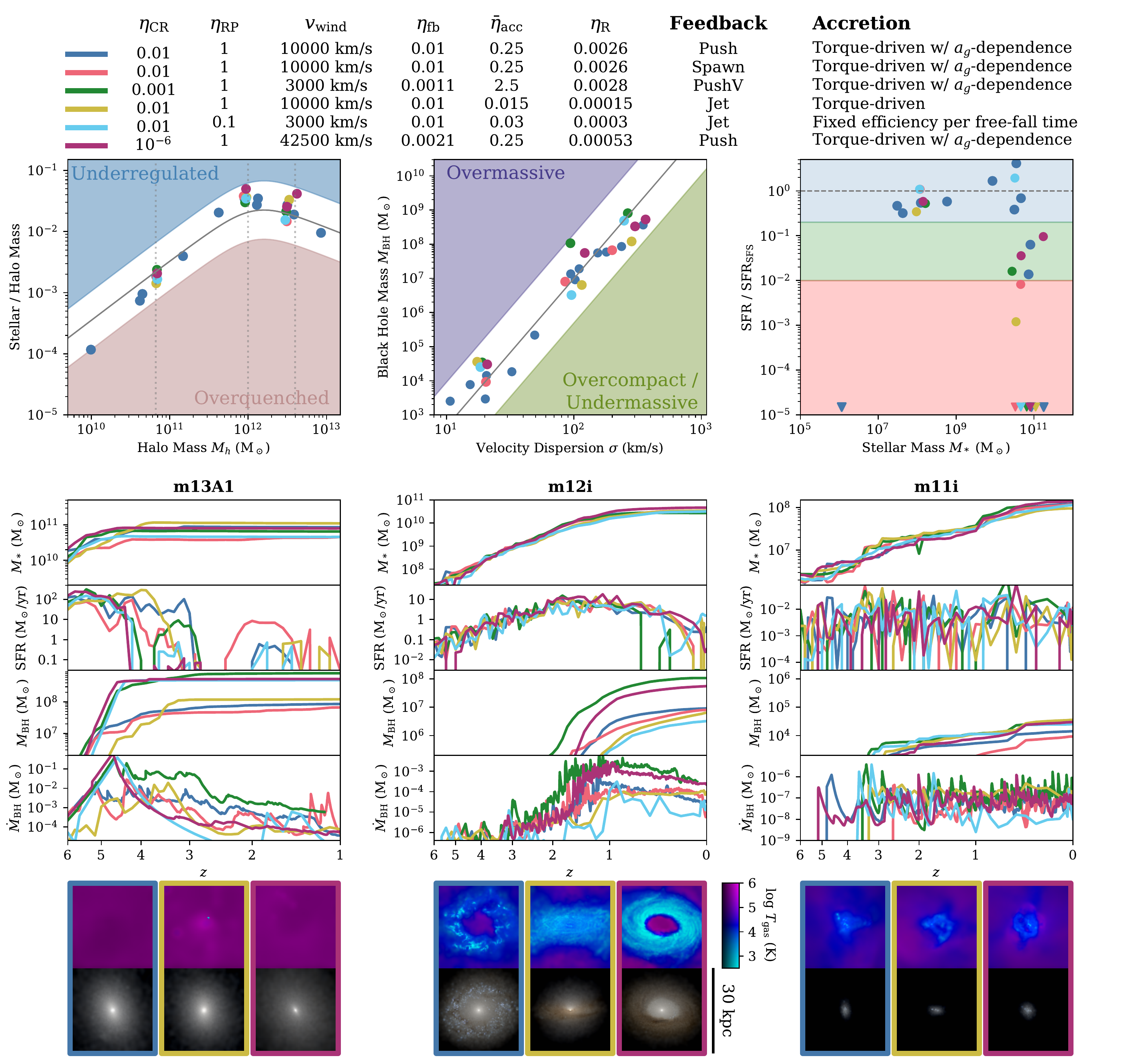}
  \caption{Several sets of model parameterizations which reasonably follow expectations for scaling relations across all masses.  Panels are defined analogously to Figure~\ref{fig:varietysets}.  This figure includes several versions of a ``fiducial" model (blue and pink), a few other variations utilizing different accretion and feedback models which achieve a similar balance, as well as a model with no cosmic ray feedback and very fast winds (purple).}
  \label{fig:greatsets}
\end{figure*}

We find that achieving appropriate regulation of star formation therefore requires: (i) a balance between accretion and feedback efficiency that regulates BH growth, and (ii) a balance between the amount and responsiveness of feedback that regulates star formation.  For a given pair of accretion and feedback models, the parameter space satisfying each of these requirements may not overlap.  Moreover, even if they do overlap at one halo mass, they may not overlap at another.  This may indicate that different models or parameterizations are more physically suitable for different halo masses.  Indeed, we find that for {\it most} combinations of accretion and feedback implementations, there is no way to parameterize them to perform well across halo mass.  Nevertheless, we do find several models within our suite which could strike this balance and perform reasonably well at regulating both BH growth and star formation across halo mass, shown in Figure~\ref{fig:greatsets} and discussed in the following section.  

\section{Models successful across halo mass}
\label{sec:goodmodels}

Although the conditions for successfully regulating both BH growth and star formation across halo mass are complex (as described in the previous two sections), there are a variety of models in our suite which are able to achieve it nonetheless.  Encouragingly, many of these ``most successful" models are also among the most physically reasonable parameter choices (e.g., with energetics in the different AGN feedback channels broadly consistent with theoretical expectations and/or observational constraints), which need not have been the case {\it a priori}.

We show six of these successful models in Figure~\ref{fig:greatsets}.  Each model has been run in at least three different halos (m11i, m12i, m13A1, and possibly others) and produced galaxy properties which lie reasonably close to the SMHM relation and $M-\sigma$ relation in all of them.  In addition, they all quench star formation in at least one massive galaxy without quenching lower-mass galaxies.\footnote{One possible exception is the leftmost blue point -- this represents m10q, a very low-mass system.  However, star formation in these systems is sporadic and so the apparent quenching may simply reflect a momentary fluctuation.}  The massive halo m13A1, whose evolution is shown in the left column, is quenched by the end of the simulation in all cases shown.  However, the moment of quenching (when the star formation rate drops by a factor $>100$) does not take place at the same time for all the models.  This variety in quenching times indicates that quenching is unlikely to be universally precipitated by an event such as a galaxy merger or interaction that would occur at the same time in each simulation, but can have different physical causes in different cases.

The first model, shown in blue, is identical to the blue model shown in Figure~\ref{fig:varietysets} and could be considered ``fiducial."  All parameter choices ($\eta_{\rm CR}$ = 0.01, $\eta_{\rm RP}$ = 1, $v_{\rm wind}$ = 10,000 km/s, and $\bar{\eta}_{\rm acc}$ = 0.25) are well within the ranges permitted by observations and physically expected by theoretical models.  This is the most well-tested of all the models shown, with 10 halos in total which all lie near the known scaling relations.  We note that the particular numerical choice of how the feedback is implemented does not appear to have a strong effect on whether these broad scaling relations are satisfied: the same parameterization with the particle-spawning feedback implementation (pink) also produces good results, as does a variation utilizing collimated jet feedback (not shown in this Figure).  

This ``fiducial" parameterization is not the only one that can produce good results, however -- the third model shown in green represents a variation with weaker feedback, but higher accretion efficiency.  This model is equally effective at achieving appropriate quenching behavior (perhaps because the responsiveness parameter $\bar{\eta}_{\rm acc}\eta_{\rm fb}$ is roughly conserved), although the BH mass in m12i is on the high end due to the higher accretion efficiency.  This model again uses a slightly different feedback implementation from the previous two (pushV vs. push and spawn), indicating that it is possible to achieve good quenching behavior with a variety of numerical choices.

\begin{figure*}
  \centering
  \includegraphics[width=1.8\columnwidth]{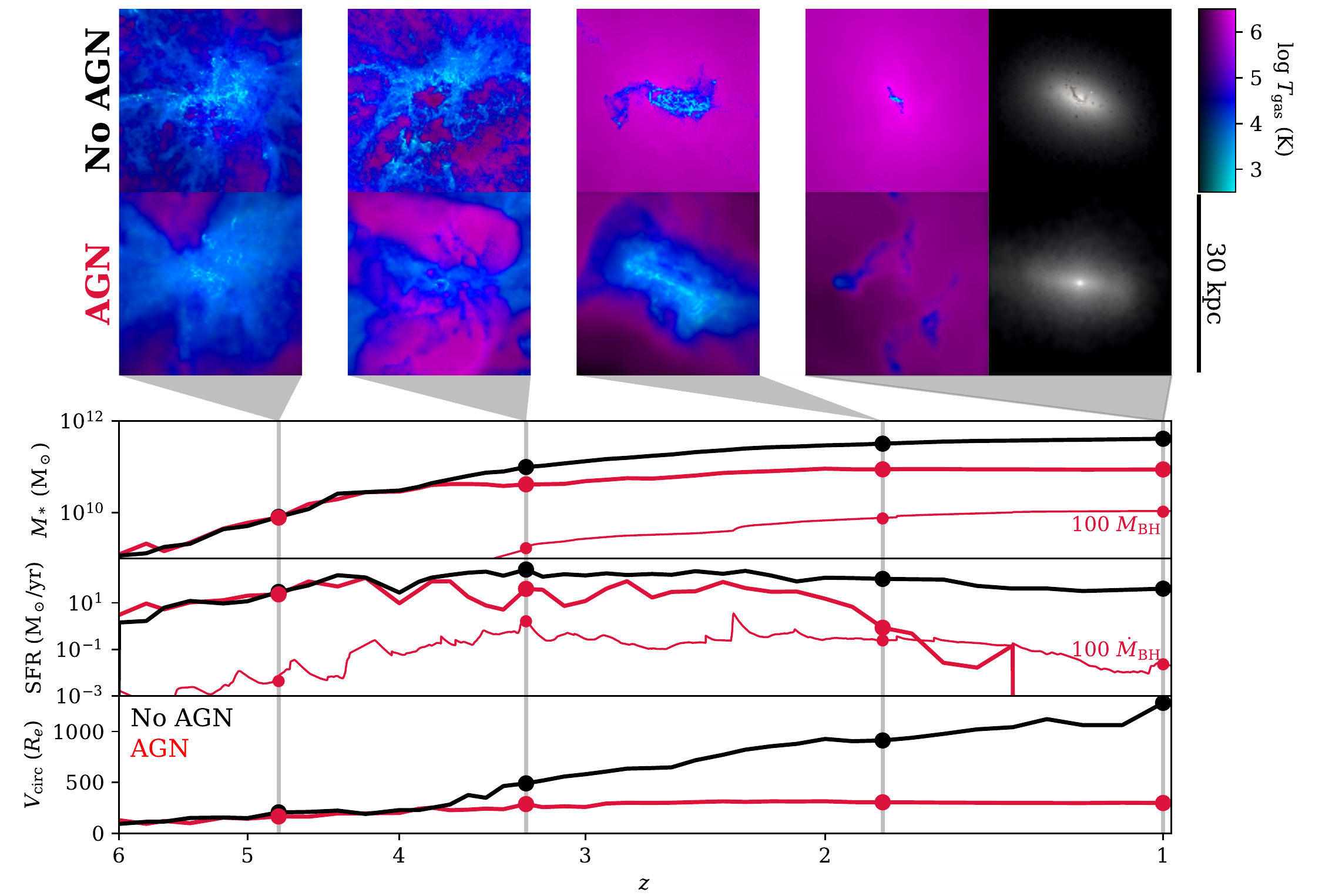}
  \caption{Comparison of a simulated massive galaxy (m13A2) with (red, bottom images) and without (black, top images) AGN physics.  (The AGN model used in this particular simulation is a version of the ``fiducial" parameterization shown in blue in Figure~\ref{fig:greatsets} with a jet feedback implementation.)  The panels in the bottom half of the figure show the redshift evolution of stellar and black hole mass (top), growth rates (middle), and the circular velocity at the effective radius (bottom).  Feedback from the SMBH reduces gas density in and around the galaxy, slowing and eventually halting star formation and preventing the formation of an extremely dense stellar nucleus.}
  \label{fig:massivegal}
\end{figure*}

The choice of feedback deposition method is not completely neutral, however -- as mentioned in previous sections, it can change the detailed effects on the galaxy and modify the balance of accretion and feedback parameters.  The fourth successful model in Figure~\ref{fig:greatsets}, shown in gold, is very similar to the gold model from Figure~\ref{fig:varietysets} which failed to quench massive galaxies.  The two models utilize the same accretion prescription and very similar parameterizations.  The main difference between the two is that the model which failed to quench uses a ``push" feedback implementation, while the successful model uses a "jet" implementation.  In this particular case, the collimation of the feedback may have permitted the BH to accrete at a high rate for a longer period of time without disrupting the inflow, resulting in a stronger overall burst of feedback and producing a quenching event which is absent in the isotropic pushing example.  Thus while the choice of feedback implementation does not drive the most important qualitative trends shown in this paper, it can affect the ``best" parameterization in detail.

The choice of {\it accretion} model is also important to whether a ``solution" can be found across halo mass.  In particular, accretion models that utilize a $a_g$-dependent prefactor to account for unresolved stellar winds (described in Section~\ref{ssec:accretion}) more easily permit good behavior across halo mass because they are less likely to allow BH overgrowth in dwarf galaxies.  Thus, the majority of the models shown in Figure~\ref{fig:greatsets} employ such an accretion model.  It is not, however, necessarily impossible to achieve this balance with other accretion models.  The fourth (gold) and fifth (cyan) models shown utilize unmodified models for torque-driven accretion and fixed accretion efficiency per free-fall time respectively.  These use similar feedback parameterizations to the fiducial model and a jet feedback implementation.  In order to avoid overgrowth of the BH in the dwarf galaxy, however, they require much lower effective accretion efficiencies, which leads to slower BH growth and somewhat undermassive BHs in the more massive systems.  

The final model, shown in purple, is an example of the parameter choices required to ``make up for" the absence of cosmic rays if that feedback channel is neglected.  Suppression of star formation in massive galaxies is still possible in this case, but requires the invocation of relatively extreme parameter values in the mechanical or radiative feedback channels.  In the example shown, the winds are launched at 42,500 km/s = 0.14$c$ at the scale of the BH kernel -- that is, they are presumed to be moving at 0.14$c$ even after traveling $\sim10$ pc from the BH.  While not impossible, this scenario is physically implausible in the sense that velocities this large are not likely to be common at this distance from the black hole.  For the radiative feedback channel, a similarly-implausible value of $\eta_{\rm RP}$ = 100 is required in order to suppress star formation in massive galaxies in the absence of cosmic ray feedback.  Both $v_{\rm wind} = 42500$ km/s and $\eta_{\rm RP} = 100$ correspond to energy efficiencies of approximately 0.01, while cosmic rays can effect quenching in massive galaxies with an energy efficiency of 0.001.  It is worth noting, however, that although the {\it energy} efficiencies required for quenching differ by up to a factor of 10 between the channels, this corresponds to a much greater difference in {\it momentum}-loading -- implying that quenching occurs through a mode which is roughly energy-conserving rather than momentum-conserving.  These examples further serve to demonstrate that there is not a {\it unique} implementation of BH physics that is capable of achieving reasonable scaling behavior across halo mass; however some implementations are more physically sensible than others, and all must satisfy the various balances discussed in the previous section.

The images at the bottom of Figure~\ref{fig:greatsets} show the gas and stars in halos m13A1, m12i, and m11i at the final snapshot of each simulation, for three variations of SMBH physics (indicated by the border color).  To compare these images to the runs without any AGN physics, see Figure~\ref{fig:controlsims}.  Within this ``successful" model parameter space, we can see similarities and differences of the outcomes across halo mass.  For m13A1, all three examples show that the gas around the galaxy is hot and much more diffuse than in the no-AGN case, indicating that the gas has been evacuated or otherwise not permitted to condense at the center.  These images show no traces of cold gas and therefore no ongoing star formation. (It is common, but not universal, for the successful models to create uniformly hot, diffuse halos around massive quenched galaxies - see Figure~\ref{fig:massivegal} for an example where some cool gas does remain in the halo). Conversely, in the case of the dwarf galaxy m11i, all three examples show cold gas and ongoing star formation, despite visible outbursts and outflows of hot gas.  The highest degree of morphological variety is to be seen in the center panels with m12i, the Milky-Way-like galaxy.  Although some cold gas is retained by the galaxy in each of these examples, the distribution of that gas differs.  The model in the central panels, which utilized jetted SMBH feedback, has a hot bipolar outflow.  The right panels, whose model includes very fast isotropic winds, show that the AGN feedback has carved out a hole in the center of the galaxy and appears to be on the verge of removing the disk entirely.  The left panels, which employ the fiducial SMBH model, also show a (less-intense) hole in the cold gas at the center of the system, but which is surrounded by dense knots of ongoing star formation.  It is clear that although the broad scaling relations are not affected by these variations in implementation, they are still likely to produce different effects on more detailed galaxy properties, a topic which will be explored in future work.

A closer time-evolution view of one of the massive galaxy simulations utilizing a well-behaved AGN feedback model is shown in Figure~\ref{fig:massivegal}.  This particular example shows halo m13A2, with a jetted version of the fiducial AGN feedback model mentioned earlier in this section.  Red lines show the time evolution of stellar mass (top), star formation rate and black hole accretion rate (middle), and circular velocity at the effective radius (bottom).  Images of selected snapshots (indicated with points and vertical grey lines) are shown in the row above.  For comparison, the no-AGN version of this halo is shown in black and in the first row of images.  In this example, the feedback from the SMBH begins to affect the star formation rate in the galaxy shortly after $z=4$.  Hot outflows generated by the feedback are already visible in the image at $z=3.2$, along with the reduction of cold dense structure in the gas which depresses the overall rate of star formation.  Star formation continues at a reduced rate with the disk somewhat inflated by ongoing feedback, but starts to drop off after a burst of BH feedback around $z=2.3$ and eventually halts entirely at $z=1.4$.  The cessation of star formation is associated with the disappearance of the gaseous disk, although a few wisps of cooler gas (which are not dense or cold enough for significant star formation) remain in the vicinity of the galaxy outskirts at $z=1$.  

\begin{figure}
  \centering
  \includegraphics[width=\columnwidth]{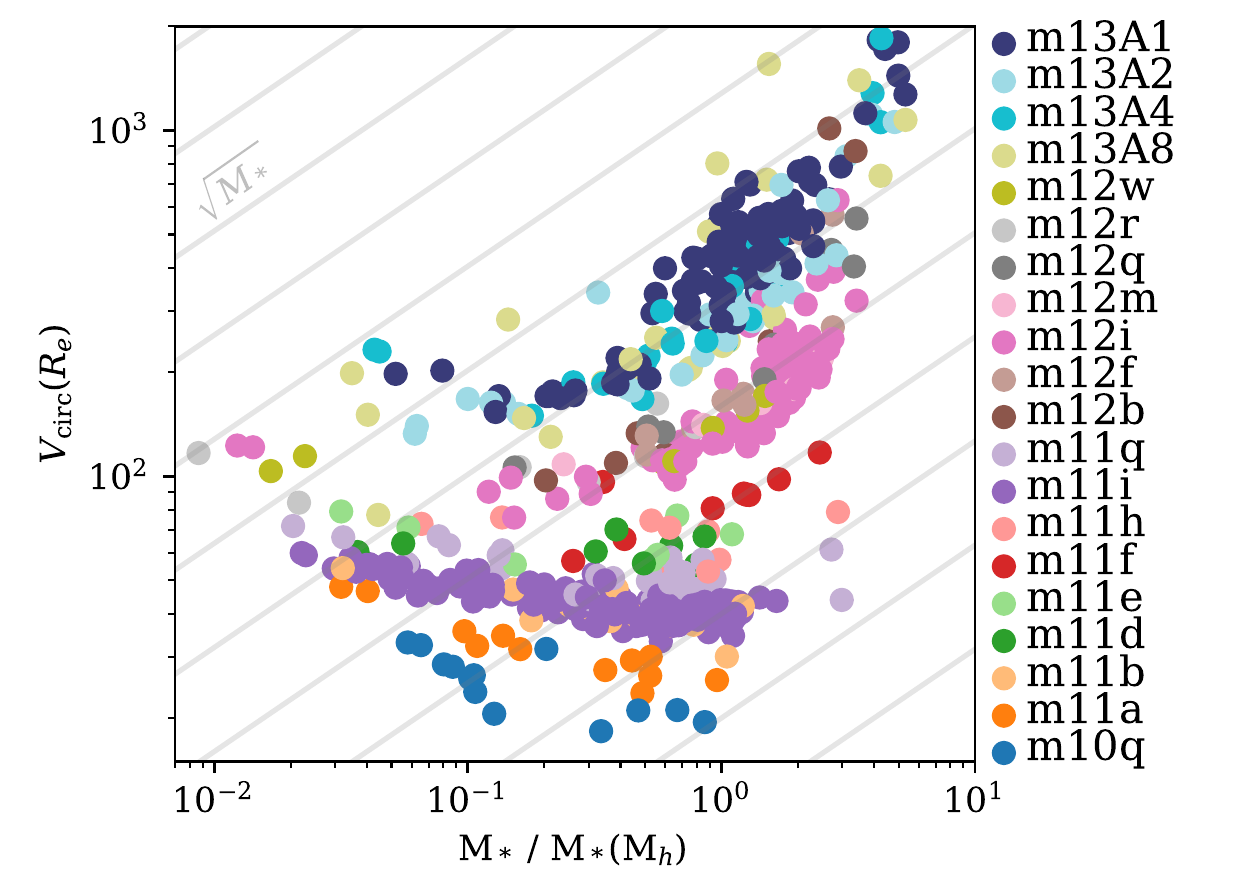}
  \caption{The relationship between a galaxy's circular velocity at the effective radius ($V_{\rm circ}(R_e)$, y-axis) and the regulation of star formation by SMBH feedback ($M_*/\langle M_* (M_h) \rangle$, x-axis).  As in Figures~\ref{fig:allsims} and \ref{fig:bhmstar}, color indicates halo identity.  For halos which lay well above the stellar mass - halo mass relation in the absence of any SMBH physics, suppressing stellar mass growth with AGN feedback significantly drives down their (previously unrealistic) circular velocities.  In this regime, the relationship is steeper than $\sqrt{M_*}$ (grey lines), indicating that the suppression of star formation is preferentially occurring in galaxy centers.}
  \label{fig:vcirc}
\end{figure}

Notably, throughout the period of time when the star formation is suppressed (from $z=4$ onward), the circular velocity remains relatively steady in the AGN run, rather than continuously increasing as it does in the no-AGN run.  In the run without SMBH physics, a dense central stellar nucleus steadily builds up over time, leading to extremely high central densities, small stellar sizes, and fast rotation curves.  As noted by \citet{Parsotan2020} and \citet{Wellons2020}, these extreme sizes and rotational velocities are inconsistent with known observational scaling relations (i.e., the size-mass and Tully-Fisher relations).  The inclusion of a reasonable model for AGN physics appears to alleviate these observational tensions, as the presence of an active BH prevents intense nuclear star formation and prevents the growth of an over-dense stellar nucleus.

The reduction of central stellar density and circular velocity by AGN feedback is a common outcome throughout the simulation suite, especially for those halos which lie well above the SMHM relation in the absence of effective SMBH feedback.  This trend can be seen in Figure~\ref{fig:vcirc} which relates circular velocity $V_{\rm circ} = \sqrt{GM(<R_e)/R_e}$ to the amount of star formation regulation in the halo, $M_*/\langle M_* (M_h)\rangle$ at the final snapshot of every simulation in the suite.  In simulations of massive galaxies where the star formation is not well-regulated (i.e., on the rightmost end of the figure), the circular velocities can reach unrealistically high values of 500-2000 km/s \citep[compare e.g. with observations of massive galaxies in][]{veale18:vel.disp}.  As the star formation is more effectively suppressed by SMBH feedback, however, the circular velocity drops significantly.  In this regime, $V_{\rm circ}$ falls faster than $\sqrt{M_*}$, indicating that the suppression of star formation is happening preferentially within $R_e$ rather than being spread evenly throughout the galaxy.  At lower masses, the relationship becomes shallower as the stellar mass no longer fully dominates the gravitational potential inside the effective radius.

Although several of the behaviors shown in the example in Figure~\ref{fig:massivegal} (e.g., the prevention of the formation of a dense stellar nucleus and the significant reduction in central gas density in the halo) do frequently appear among the simulations in our AGN suite, not all aspects of it are representative.  In particular, we note that the relatively ``slow" quenching shown in this example does not reflect the suite as a whole -- the suite contains runs with a wide variety of quenching timescales, some of which are much more rapid than this particular example would imply.  (See the SFR panels in Figures~\ref{fig:varietysets} and \ref{fig:greatsets} for several more examples of quenching times and timescales.)

The set of balances between accretion, feedback, black hole growth, and star formation required to reproduce observed galaxy quenching behavior is complex, and cannot necessarily be satisfied by an arbitrary implementation of BH accretion and feedback simply by finding the ``correct" adjustments to the model parameters.  Nevertheless, we find within our suite several different combinations of accretion and feedback models that do permit parameterizations which successfully reproduce broad scaling relations across halo mass.  Moreover, we find that these successful AGN models can also affect other properties of massive galaxies, such as central densities and rotation curves, in ways that tend to bring them more in line with observed systems.   

\section{Conclusions}
\label{sec:conclusions}

To better understand the physical relationship between supermassive black hole (SMBH) growth and galaxy quenching, we have run and analyzed a suite of approximately 500 zoom-in simulations of galaxy formation which include a variety of models for SMBH accretion and feedback.  The simulations range across three decades in halo mass from present-day dwarf galaxies ($\sim 10^{10}$~\Msun~at $z=0$) to high-redshift massive galaxies ($\sim 10^{13}$~\Msun~at $z=1$).  The suite explores variations in accretion model, accretion efficiency, feedback injection scheme, and the efficiency of AGN feedback in three independent channels: (i) radiation (including photo-heating, ionization, Compton heating, and radiation pressure), (ii) mechanical winds (including thermal, kinetic, and magnetic components), and  (iii) cosmic rays.

Analogous simulations which do not include SMBH physics (i.e., the standard FIRE-2 simulations which appear in numerous other works) have been shown to be highly successful at reproducing observed galaxies up to about Milky Way mass.  At higher masses ($M_{\rm halo} \gtrsim 10^{12}$~\Msun), however, simulated galaxies differ from observations in several significant ways: they are overly compact, with extremely high rotational velocities, and most importantly, they universally fail to quench (stop forming stars, see Figure~\ref{fig:controlsims}).  The inclusion of SMBH physics in massive galaxy simulations can introduce quenching through the AGN feedback processes.  We examine our simulation suite to determine what SMBH model properties lead to good quenching behavior in massive galaxies, while avoiding overquenching in less-massive systems.  We use three well-known galaxy scaling relations to assess the behavior: the stellar mass - halo mass relation, the $M-\sigma$ relation between black hole mass and galaxy velocity dispersion, and the star formation rate - stellar mass relation.

We find a variety of outcomes within the suite.  Some models still fail to quench massive galaxies despite the inclusion of AGN feedback, many overquench star formation (especially in dwarf galaxies), and others are able to appropriately regulate star formation.  We find that the regulation of star formation is closely related to the regulation of black hole growth: overly-massive black holes tend to overquench their host galaxies, while undermassive black holes often fail to quench massive systems (Figure~\ref{fig:bhmstar}).  

Within our suite, we find several general trends between accretion efficiency, feedback efficiencies, black hole growth, and stellar mass growth which are relatively independent of numerical implementation.  Though both the accretion and feedback efficiencies play a role in determining black hole mass, in massive systems (Milky-Way-mass and above) the rate of BH growth is most directly affected by the choice of accretion efficiency.  In dwarf galaxies, however, the BH growth is most sensitive to the choice of feedback efficiency: models with stronger feedback produce smaller black holes, as the winds suppress further accretion regardless of accretion efficiency.  (Figure~\ref{fig:bhreg})

The degree to which star formation is suppressed in our simulated galaxies depends on (i) the total amount of AGN feedback energy released, (ii) the responsiveness of the feedback model to changes in the gas supply, and (iii) the halo mass of the system (Figure~\ref{fig:sfreg}).  At low halo masses, any SMBH model that produces enough feedback energy to unbind the baryons in the halo overquenches the galaxy in our simulations.  In more massive systems, this is a necessary but not sufficient condition for quenching: the degree to which star formation is suppressed depends not only on the total amount of energy released in feedback, but on how quickly and forcefully that energy was released.  We quantify this using the product of the accretion and feedback efficiencies of the SMBH model, and term that quantity ``responsiveness" (Eqn.~\ref{eqn:responsiveness}).  Models with high responsiveness react impulsively and powerfully to changes in the BH environment, and in turn give the galaxy itself less time to respond to that influx of energy.  In massive galaxies, we find that SMBH models with higher responsiveness more strongly suppress star formation in the galaxy through their explosive behavior, while low-responsiveness models which are more steady are less able to disrupt the system.

These broad trends with accretion and feedback efficiency hold regardless of the details of the numerical implementation.  However, we do find noticeable effects from certain other modeling choices.  When comparing the three different AGN feedback channels we employed (radiation, mechanical winds, and cosmic rays), we find that the cosmic rays are the most efficient at regulating star formation.  Star formation can be significantly suppressed by AGN models with only a modest value for the efficiency of cosmic ray production, but achieving quenching through mechanical winds or radiation alone requires the invocation of very high feedback efficiencies through very fast winds or a large artificial boost to radiation pressure.  These results are consistent with studies of AGN jets at the cluster scale by \citet{Su2019, Su2020}, who find that cosmic ray jets are more effective at suppressing cooling flows than thermal heating or momentum injection, and produce better agreement with X-ray cluster observables.

One important caveat for cosmic rays in particular is that we assume a constant scattering rate/diffusion coefficient. While this coefficient is quite well-constrained in the local ISM (from detailed Solar system observations) and the typical ISM of nearby galaxies (from $\gamma$-ray observations), the true physical scattering rates almost certainly vary in different environments and galaxies (e.g. dense galactic nuclei, or the low-density CGM, where CRs influence cooling onto galaxies). Different models calibrated to reproduce the same ISM data but making different extrapolations for the effective CR transport coefficients in the CGM can significantly influence the effect of CRs on star formation even in the absence of AGN \citep{hopkins:2020.cr.transport.model.fx.galform,hopkins:2021.sc.et.models.incompatible.obs}. So it is especially important to consider in future work the observational consequences of different CR transport physics in different environments.

We find that the choice of accretion model is important to achieving good mass-dependent behavior, and that many accretion models produce qualitatively incorrect behavior which cannot be corrected simply by adjusting the parameterization (including many popular variations of Bondi accretion).  Many of the most ``successful" models in our suite employ an accretion model which includes a correction for mass loss due to unresolved stellar feedback.  Finally, we found that varying among our four numerical/geometric injection schemes for feedback produced subdominant, but not completely negligible differences -- for example, we find in some cases that highly-collimated feedback may take longer to self-regulate the BH growth than isotropic feedback does.

Given the complexity and halo-mass-dependence of the interplay between the numerical implementation, accretion and feedback efficiencies, and the outcomes for BH and galaxy growth, most of the implementations for SMBH physics explored in our suite cannot necessarily be ``tuned" to achieve appropriate quenching behavior across entire the range of halo mass we tested.  Many of the strongest constraints on the models come from the inclusion of dwarf galaxies ($M_h < 10^{11}$ \Msun) in our sample, which are not well resolved in most previous galaxy formation simulations including AGN feedback.  Nevertheless, we did identify several models which were able to satisfy these balances and generally reproduce the known scaling relations (Figure~\ref{fig:greatsets}).  Of these, many have very reasonable physical parameterizations, with cosmic ray feedback efficiency $\eta_{\rm CR} \sim 0.001-0.01$, mechanical wind velocity $v_{\rm wind} \sim$ 3,000 - 10,000 km/s, and radiative feedback which is not artificially boosted ($\eta_{\rm RP} \sim 1$).\footnote{In terms of computational expense, adding these AGN feedback channels, even with our high-resolution ``spawn'' routines, generally incurs little additional expense compared to models with identical physics (radiation, magnetic fields, cosmic rays) but no AGN feedback. In fact, because of the reduced stellar masses and central densities, the net CPU cost of AGN feedback runs is often greatly reduced in massive halos compared to no-feedback analogues. It is, however, the case that some physics (for stars and/or AGN) are significantly more expensive: simulations with cosmic rays, for example, can be $\sim 5-10$ times more expensive (all else equal, in e.g.\ dwarf or Milky Way-mass halos), compared to simulations without cosmic rays (primarily owing to the shorter timesteps required for explicit CR transport).}  We expect that the exact parameter values for the ``best" models will be somewhat sensitive to changes in other aspects of the simulations such as resolution or details of the stellar physics, but the general trends described here are likely robust.  It is encouraging that SMBH models exist which can reproduce basic scaling relations across halo mass down to the dwarf galaxy scale.  It is also possible, however, that a more complex, halo-mass-dependent model would be a better physical representation of how SMBHs interact with their host galaxies as the properties of the ISM and CGM change across mass scales.

In this study, we have focused on very broad galaxy properties (such as stellar mass, star formation rate, and black hole mass) to assess the performance of SMBH models within the context of galaxy quenching.  The modeling choices we explored here will certainly also affect more detailed galaxy properties (such as morphology, or the effects on the ISM and CGM observables).  Future work on these topics may therefore serve to further distinguish between the most plausible implementations and parameter values. 
Simulations like the ones analyzed in this paper, in which different stellar and AGN feedback channels are explicitly modeled, can also be used to analyze in greater depth exactly how the different feedback mechanisms act on different physical scales.

\section*{Acknowledgements}

SW is supported by an NSF Astronomy and Astrophysics Postdoctoral Fellowship under award AST2001905.  
CAFG was supported by NSF through grants AST-1715216, AST-2108230,  and CAREER award AST-1652522; by NASA through grant 17-ATP17-0067; by STScI through grant HST-AR-16124.001-A; and by the Research Corporation for Science Advancement through a Cottrell Scholar Award and a Scialog Award. 
EQ was supported in part by a Simons Investigator grant from the Simons Foundation and NSF AST grant 2107872.
DAA was supported in part by NSF grants AST-2009687 and AST-2108944.
RF acknowledges financial support from the Swiss National Science Foundation (grant no 194814).
DK was supported by NSF through grants AST-1715101 and AST2108314.
AW received support from: the NSF via CAREER award AST-2045928 and grant AST-2107772; NASA ATP grant 80NSSC20K0513; HST grants AR-15809 and GO-15902 from STScI.
This work was performed in part at Aspen Center for Physics, which is supported by National Science Foundation grant PHY-1607611.  
The data used in this work were, in part, hosted on facilities supported by the Scientific Computing Core at the Flatiron Institute, a division of the Simons Foundation.

\section*{Data Availability}

The data supporting the plots within this article are available on reasonable request to the corresponding author. A public version of the GIZMO code is available at http://www.tapir.caltech.edu/$\sim$phopkins/Site/GIZMO.html. 
Additional data including simulation snapshots, initial conditions, and derived data products are available at https://fire.northwestern.edu/data/.

\bibliography{biblio.bib}

\bsp	
\label{lastpage}
\end{document}